\documentclass[12pt,prd,showpacs,tightenlines,nofootinbib]{revtex4}
\usepackage{bm}
\usepackage{graphics}
\usepackage{rotating}
\usepackage{epsfig}
\begin{document}
\title{Rare $B_s$ decays in the
  relativistic quark model } 

\author{R. N. Faustov}
\author{V. O. Galkin}
\affiliation{Dorodnicyn Computing Centre, Russian Academy of Sciences,
  Vavilov Str. 40, 119333 Moscow, Russia}

\begin{abstract}
The branchings fractions of the rare $B_s$ decays are calculated in the
framework of the QCD-motivated relativistic quark model. The form
factors of the weak $B_s$ transitions to light mesons are
expressed through the overlap integral of the initial and final meson
wave functions in the whole accessible kinematical range. Explicit
determination of the momentum transfer dependence of the form factors
without additional model assumptions and extrapolations significantly
improve the reliability of the obtained results. The approximate
analytical form of the form factors is given in order to simplify the
comparison with other predictions and experiment. The calculated form
factors are applied for the investigations of the rare semileptonic,
radiative and nonleptonic $B_s$ decays. The factorization
approximation is used for the description of the nonleptonic decays. All
results agree well with available experimental data. 

\end{abstract}


\maketitle

\section{Introduction}
\label{sec:int}

The rare weak decays of $B_s$ mesons are governed by the flavour
changing neutral current. In the standard model their description requires
calculation of the loop (penguin) diagrams. Thus such decays are
very sensitive to the intermediate contributions of the new particles
and interactions. The accurate theoretical evaluation and
experimental measurement of their decay rates can significantly constrain the ``new
physics'' models. 

Theoretical investigation of these rare decays is
usually based on the effective Hamiltonian in which intermediate gauge
bosons are integrated out. Application of the operator product
expansion allows one to separate short- and long-distance effects
which are assumed to factorize. The short-distance 
contributions are described by the Wilson coefficients which are
calculated perturbatively. The long-distance part is attributed to the
set of the operators, which matrix elements between initial and final
meson states are usually parametrized by the set
of the invariant form factors. The calculation of these form factors
requires application of the nonperturbative methods. Thus the
improvement of the theoretical understanding of the rare decays
requires the precise control of hadronic uncertainties. The
characteristic feature of the rare semileptonic $B_s$ decays is a
very broad kinematical range. Therefore the reliable determination of
the momentum transfer dependence of the form factors turn out to be very important. Various
theoretical approaches have been applied for the form factor
calculations. However, in most of such approaches the decay form
factors are determined in some specific kinematic point or interval and
then they are extrapolated to the whole kinematical range
using some model parametrizations or additional assumptions. Thus in
the region of the large recoil of the final meson light-cone QCD sum
rules can be applied, while the region of small recoils is accessible
to lattice QCD. Most of quark models determine form factors at the
single point of zero or maximum recoil and then the Gaussian or pole
extrapolations are applied. Therefore  the
calculation of the form factors in the whole kinematical range in the
framework of the same approach without additional assumptions or/and
extrapolations is highly important.

From the experimental side significant progress has been achieved in last
years \cite{pdg,lhcbbr,bellebs,lhcbrad,lhcbnl,lhcbnl2}. Not only
several rare $B_s$ decays have been observed for the
first time and their branching fractions were measured, but also first
attempt have been made for measuring the differential branching fraction and
longitudinal polarization fraction of the $\varphi$ meson in the rare
semileptonic $B_s\to\varphi\mu^+\mu^-$ decay \cite{lhcbbr}.

In this paper we calculate the rare weak $B_s$ transition form factors
and decay rates in
the framework of the relativistic quark model based on the
quasipotential approach in quantum chromoynamics (QCD). It was
previously successfully applied for evaluating the charmfull and charmless
semileptonic $B_s$ decay form factors \cite{bsdecay,bsld}. Rare
semileptonic decays of $B$ and $B_c$ mesons were studied by us in
Ref.~\cite{brare}. This model
consistently takes relativistic effects into account. It allows us to
express the form factors of the transition matrix elements through the overlap integrals of the
meson wave functions. Such expressions are valid in the whole
kinematical range and thus do not require additional assumptions or
extrapolations. The meson wave functions are known in our model from
the previous mass spectra calculations \cite{hlm,lm}. On the basis
of the form factors we then consider the rare semileptonic, radiative
and nonleptonic $B_s$ decays.

The paper is organized as follows. First, in Sec.~\ref{rqm} we briefly describe
our relativistic quark model paying special attention to the method of
the calculation of the transition matrix element of the weak current.
In Sec.~\ref{sec:ffbsphi} using
this approach we calculate the form factors
of the rear weak $B_s\to\eta(\varphi)$ transitions. The momentum
transfer dependence of the form factors is determined explicitly and
the approximate parametrizations of the form factors are given.  In
Sec.~\ref{sec:rard} these form factors are used for consideration
of the rare semileptonic $B_s\to \eta(\varphi)l^+l^-$ decays.  The
$B_s\to \eta(\varphi)\nu\bar\nu$ decays are discussed in
Sec.~\ref{sec:bsnu}. Section~\ref{sec:rrad} contains results for the rare
radiative decays. 
In Sec.~\ref{sec:nonl} rare nonleptonic $B_s$ decays to a charmonium
state and a light meson as well as to two light mesons are calculated
 in the framework of the factorization approximation.
All obtained results are confronted
with previous predictions and available experimental
data. Section~\ref{sec:concl} contains our conclusions.           

\section{Relativistic quark model}  
\label{rqm}

\subsection{Relativistic wave equation and the quasipotential of the quark-antiquark interaction}
\label{sec:qe}

For further calculations we use the relativistic quark model based on the
quasipotential approach in quantum chromodynamics (QCD). In our model hadrons are
considered as the bound states of constituent quarks which are described by the
single-time wave functions satisfying the
three-dimensional relativistically invariant Schr\"odinger-like
equation with the QCD-motivated quark-antiquark potential \cite{mass}  
\begin{equation}
\label{quas}
{\left(\frac{b^2(M)}{2\mu_{R}}-\frac{{\bf
p}^2}{2\mu_{R}}\right)\Psi_{M}({\bf p})} =\int\frac{d^3 q}{(2\pi)^3}
 V({\bf p,q};M)\Psi_{M}({\bf q}),
\end{equation}
where the relativistic reduced mass is
\begin{equation}
\mu_{R}=\frac{M^4-(m^2_1-m^2_2)^2}{4M^3},
\end{equation}
 $M$ is the meson mass, $m_{1,2}$ are the quark masses,
and ${\bf p}$ is their relative momentum.  
In the center of mass system the relative momentum squared on mass shell 
$b^2(M)$ is expressed through the meson and quark masses:
\begin{equation}
{b^2(M) }
=\frac{[M^2-(m_1+m_2)^2][M^2-(m_1-m_2)^2]}{4M^2}.
\end{equation}
The quark-antiquark quasipotential $V({\bf
  p,q};M)$ is assumed to be the sum of the perturbative one-gluon
exchange term and the nonperturbative confining part \cite{mass}
  \begin{equation}
\label{qpot}
V({\bf p,q};M)=\bar{u}_1(p)\bar{u}_2(-p){\mathcal V}({\bf p}, {\bf
q};M)u_1(q)u_2(-q), 
\end{equation}
with
$${\mathcal V}({\bf p},{\bf q};M)=\frac{4}{3}\alpha_sD_{ \mu\nu}({\bf
k})\gamma_1^{\mu}\gamma_2^{\nu}
+V^V_{\rm conf}({\bf k})\Gamma_1^{\mu}({\bf k})
\Gamma_{2;\mu}({\bf k})+V^S_{\rm conf}({\bf k}),\qquad {\bf k=p-q},$$
where $\alpha_s$ is the QCD coupling constant, $D_{\mu\nu}$ is the
gluon propagator in the Coulomb gauge, and $\gamma_{\mu}$ and $u(p)$ are  the Dirac matrices and
spinors, respectively. The confining part consists from the  Lorentz scalar and vector linearly
rising interactions which in the nonrelativistic limit reduce to
\begin{equation}
\label{nr}
V_{\rm conf}(r)=V_{\rm conf}^S(r)+V_{\rm conf}^V(r)=Ar+B,
\end{equation}
with
\begin{equation}
\label{vlin}
V_{\rm conf}^V(r)=(1-\varepsilon)(Ar+B),\qquad 
V_{\rm conf}^S(r) =\varepsilon (Ar+B),
\end{equation}
where $\varepsilon$ is the mixing coefficient. Its value  $\varepsilon=-1$
has been determined from the comparison of the heavy quark expansion
for the semileptonic $B\to D$ decays in our model
\cite{fg} with model-independent predictions of heavy quark effective
theory and from the consideration of charmonium radiative decays
\cite{mass}. Note that  in the nonrelativistic limit the quasipotential
(\ref{qpot}) reproduces the well-known Cornell potential. Therefore
this quasipotential provides its relativistic generalization. 

The long-range vector vertex 
\begin{equation}
\label{kappa}
\Gamma_{\mu}({\bf k})=\gamma_{\mu}+
\frac{i\kappa}{2m}\sigma_{\mu\nu}k^{\nu}
\end{equation}
contains the Pauli
term with anomalous chromomagnetic quark moment $\kappa$. The  value
$\kappa=-1$ is
fixed in our model from the analysis of the fine splitting of heavy quarkonia ${
}^3P_J$- states \cite{mass} and  the heavy quark expansion for semileptonic
decays of heavy mesons \cite{fg} and baryons \cite{sbar} and 
enables vanishing of the spin-dependent chromomagnetic interaction,
proportional to $(1+\kappa)$, in
accord with the flux tube model. 

Other parameters of our model  were determined from the previous analysis of
meson spectroscopy  \cite{mass}. The
constituent quark masses
are $m_b=4.88$ GeV, $m_c=1.55$ GeV, $m_s=0.5$ GeV, $m_{u,d}=0.33$ GeV and
the parameters of the linear potential are $A=0.18$ GeV$^2$ and
$B=-0.30$ GeV.  

\subsection{Matrix element of the weak current between meson states}
\label{sec:mem}

The calculation of the branching fractions of rear weak decays requires
evaluation of the transition matrix elements of the weak current $J_\mu^W$ between meson
states. In the quasipotential approach such matrix element between a $B_s$ meson with mass $M_{B_s}$ and
momentum $p_{B_s}$ and a final $f$ ($\eta,\eta'$ or $\varphi$) meson with mass $M_{f}$ and
momentum $p_{f}$ is given by  \cite{f}
\begin{equation}\label{mxet} 
\langle f(p_{f}) \vert J^W_\mu \vert B_s(p_{B_s})\rangle
=\int \frac{d^3p\, d^3q}{(2\pi )^6} \bar \Psi_{{f}\,{\bf p}_f}({\bf
p})\Gamma _\mu ({\bf p},{\bf q})\Psi_{B_s\,{\bf p}_{B_s}}({\bf q}),
\end{equation}
where $\Gamma _\mu ({\bf p},{\bf
q})$ is the two-particle vertex function and  
$\Psi_{M\,{\bf p}_M}({\bf p})$ are the
meson ($M=B_s,{f})$ wave functions projected onto the positive energy states of
quarks and boosted to the moving reference frame with momentum ${\bf
  p}_M$, and  ${\bf p},{\bf q}$ are relative quark momenta.

 \begin{figure}
  \centering
  \includegraphics{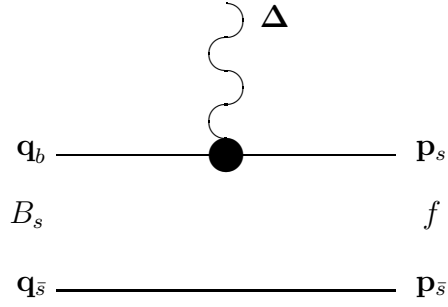}
\caption{Leading order vertex function $\Gamma^{(1)}({\bf p},{\bf q})$. \label{d1}}
\end{figure}

\begin{figure}
  \centering
  \includegraphics{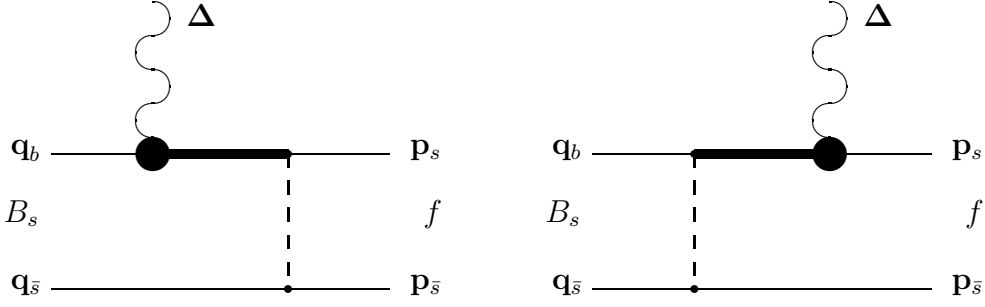}
\caption{ Subleading order vertex function $\Gamma^{(2)}({\bf p},{\bf
    q})$. Bold lines denote the negative-energy part of the quark
  propagator. Dashed lines correspond  to the exchange by the effective potential ${\cal V}$ 
(\ref{qpot}).  \label{d2}} 
\end{figure}

The vertex function  $\Gamma_\mu ({\bf
  p},{\bf q})$ contains \cite{bsdecay}
contributions both from the leading order spectator diagram (Fig.~\ref{d1})  and from
subleading order diagrams (Fig.~\ref{d2}) accounting for the contributions of the
negative-energy intermediate states. The leading order vertex function
\begin{equation} \label{gamma1}
\Gamma_\mu^{(1)}({\bf
p},{\bf q})=\bar u_{s}(p_s)J_\mu^W u_b(q_b)
(2\pi)^3\delta({\bf p}_{\bar s}-{\bf
q}_{\bar s})\end{equation}
contains the $\delta$ function which allows us to take one of the
integrals in the matrix element (\ref{mxet}) and thus to reduce it to the
standard overlap integral of meson wave functions.  The subleading order
contribution is significantly more complicated 
\begin{eqnarray}\label{gamma2} 
\Gamma_\mu^{(2)}({\bf
p},{\bf q})&=&\bar u_{s}(p_s)\bar u_s(p_{\bar s}) \Bigl\{{\cal V}({\bf p}_{\bar s}-{\bf
q}_{\bar s})\frac{\Lambda_{s}^{(-)}(k')}{ \epsilon_{s}(k')+
\epsilon_{s}(q_b)}\gamma_1^0 J_\mu^W\nonumber \\ 
& &+J_\mu^W
\frac{\Lambda_b^{(-)}(
k)}{\epsilon_b(k)+\epsilon_b(p_s)}\gamma_1^0
{\cal V}({\bf p}_{\bar s}-{\bf
q}_{\bar s})\Bigr\}u_b(q_b)
u_s(q_{\bar s}),\end{eqnarray}
where ${\bf k}={\bf p}_s-{\bf\Delta};\
{\bf k}'={\bf q}_b+{\bf\Delta};\ {\bf\Delta}={\bf
p}_{f}-{\bf p}_{B_s}$;
$$\Lambda^{(-)}(p)=\frac{\epsilon(p)-\bigl( m\gamma
^0+\gamma^0({\bm{ \gamma}{\bf p}})\bigr)}{ 2\epsilon (p)}, \qquad
\epsilon(p)=\sqrt{{\bf p}^2+m^2}.$$
It depends in a very complicated way on
the relative momenta of quarks which enter the energies of the initial heavy and final
light quarks. For the heavy quark energy the  heavy quark
expansion can be applied. For the light quark such expansion is not valid.  However, 
the final light $f$ meson possesses a large recoil momentum
($|{\bf\Delta}_{\rm  max}|=(M_{B_s}^2-M_f^2)/(2M_{B_s})\sim 2.6$~GeV),
with respect to the mean relative quark momentum $|{\bf p}|$ in the meson ($\sim 0.5$~GeV), almost in
the whole kinematical range except the small region near  
$q^2=q^2_{\rm max}$ ($|{\bf\Delta}|=0$).  
This observation allows one to neglect  $|{\bf p}|$ compared to $|{\bf\Delta}|$ in  the
light quark energy
$\epsilon_{q}(p+\Delta)\equiv\sqrt{m_{q}^2+({\bf 
p}+{\bf\Delta})^2}$, replacing it by  $\epsilon_{q}(\Delta)\equiv
\sqrt{m_{q}^2+{\bf\Delta}^2}$  in expressions for the
subleading contribution.  Such replacement removes the relative
momentum dependence in the energies of  quarks and thus permits
to perform one of the integrations in the subleading 
contribution using the quasipotential equation. Since the subleading
contributions are additionally suppressed by the ratio of the small binding energy
to the large total energy of the meson, the uncertainty introduced by such
procedure is small.  As the result the weak
decay matrix element is expressed through the usual overlap integral
of initial and final meson wave functions and its momentum dependence
can be determined in the whole accessible kinematical range without
additional assumptions. 

It is important to point out that initial and final
mesons in the considered rare decay are moving with respect to each other.
This fact should be taken into account in calculating the decay 
matrix element (\ref{mxet}). If
calculations are done in the $B_s$ meson
rest frame (${\bf p}_{B_s}=0$) then the final  meson
is moving with the recoil momentum ${\bf p}_{f}={\bf \Delta}$. In the
quasipotential approach the wave function
of the moving  meson $\Psi_{f\,{\bf\Delta}}$ is connected 
with the  wave function in the rest frame 
$\Psi_{f\,{\bf 0}}\equiv \Psi_{f}$ by the relativistic transformation \cite{f}
\begin{equation}
\label{wig}
\Psi_{f\,{\bf\Delta}}({\bf
p})=D_s^{1/2}(R_{L_{\bf\Delta}}^W)D_s^{1/2}(R_{L_{
\bf\Delta}}^W)\Psi_{f\,{\bf 0}}({\bf p}),
\end{equation}
where $R^W$ is the Wigner rotation, $L_{\bf\Delta}$ is the Lorentz boost
 to a moving reference frame and $D^{1/2}(R)$ is  
the spin rotation matrix.

\section{Form factors of the rear weak transitions of $B_s$ to $\eta(\varphi)$ mesons}
\label{sec:ffbsphi}

The matrix elements of the flavour changing neutral currents, governing rare $b\to s$ weak transitions, between initial $B_s$ meson and final 
$\eta$ or $\varphi$ mesons are usually parametrized by the following set
of the invariant form factors 
\begin{equation}
  \label{eq:pff1}
  \langle \eta(p_{\eta})|\bar s \gamma^\mu b|B_s(p_{B_s})\rangle
  =f_+(q^2)\left[p_{B_s}^\mu+ p_{\eta}^\mu-
\frac{M_{B_s}^2-M_{\eta}^2}{q^2}\ q^\mu\right]+
  f_0(q^2)\frac{M_{B_s}^2-M_{\eta}^2}{q^2}\ q^\mu,
\end{equation}
\begin{equation}
\label{eq:pff12} 
 \langle \eta(p_{\eta})|\bar s \gamma^\mu\gamma_5 b|B_s(p_{B_s})\rangle
  =0,
\end{equation}
\begin{equation}
\label{eq:pff2}
\langle \eta(p_\eta)|\bar s \sigma^{\mu\nu}q_\nu b|B_s(p_{B_s})\rangle=
\frac{if_T(q^2)}{M_{B_s}+M_\eta} [q^2(p_{B_s}^\mu+p_\eta^\mu)-(M_{B_s}^2-M_\eta^2)q^\mu],
\end{equation}
\begin{eqnarray}
  \label{eq:vff1}
  \langle {\varphi}(p_{\varphi})|\bar s \gamma^\mu b|B_s(p_{B_s})\rangle&=
  &\frac{2iV(q^2)}{M_{B_s}+M_{\varphi}} \epsilon^{\mu\nu\rho\sigma}\epsilon^*_\nu
  p_{B_s\rho} p_{{\varphi}\sigma},\\ \cr
\label{eq:vff2}
\langle {\varphi}(p_{\varphi})|\bar s \gamma^\mu\gamma_5 b|B_s(p_{B_s})\rangle&=&2M_{\varphi}
A_0(q^2)\frac{\epsilon^*\cdot q}{q^2}\ q^\mu
 +(M_{B_s}+M_{\varphi})A_1(q^2)\left(\epsilon^{*\mu}-\frac{\epsilon^*\cdot
    q}{q^2}\ q^\mu\right)\cr\cr
&&-A_2(q^2)\frac{\epsilon^*\cdot q}{M_{B_s}+M_{\varphi}}\left[p_{B_s}^\mu+
  p_{\varphi}^\mu-\frac{M_{B_s}^2-M_{\varphi}^2}{q^2}\ q^\mu\right], 
\end{eqnarray}
\begin{equation}
  \label{eq:vff3}
\langle  \varphi(p_{\varphi})|\bar s i\sigma^{\mu\nu}q_\nu b|B_s(p_{B_s})\rangle=2T_1(q^2)
\epsilon^{\mu\nu\rho\sigma} \epsilon^*_\nu p_{{\varphi}\rho} p_{{B_s}\sigma},
\end{equation}
\begin{eqnarray}
\label{eq:vff4}
\langle \varphi(p_{\varphi})|\bar s i\sigma^{\mu\nu}\gamma_5q_\nu b|B_s(p_{B_s})\rangle&=&
T_2(q^2)[(M_{B_s}^2-M_{\varphi}^2)\epsilon^{*\mu}-(\epsilon^*\cdot q)(p_{B_s}^\mu+
p_{\varphi}^\mu)]\cr\cr
&&+T_3(q^2)(\epsilon^*\cdot q)\left[q^\mu-\frac{q^2}{M_{B_s}^2-M_{\varphi}^2}
  (p_{B_s}^\mu+p_{\varphi}^\mu)\right],
\end{eqnarray}
were $q=p_{B_s}- p_{\eta(\varphi)}$ is the momentum transfer,
$M_{B_s,\eta(\varphi)}$ are the initial and final meson masses, and  $\epsilon_\mu$ is the
  polarization vector of the final vector  $\varphi$ meson.

At the maximum recoil point ($q^2=0$) these form
factors satisfy the following conditions: 
\[f_+(0)=f_0(0),\]
\[A_0(0)=\frac{M_{B_s}+M_\varphi}{2M_{\varphi}}A_1(0)
-\frac{M_{B_s}-M_{\varphi}}{2M_{\varphi}}A_2(0),\]
\[T_1(0)=T_2(0).\]

The physical pseudoscalar $\eta$ and $\eta'$ mesons are the mixtures of
$\eta_q( u \bar u+ d\bar d)$ and $\eta_s( s\bar s)$ states
\begin{eqnarray}
  \label{eq:mix}
  |\eta\rangle&=&|\eta_q\rangle\cos\phi-|\eta_s\rangle\sin\phi, \\
 |\eta'\rangle&=&|\eta_q\rangle\sin\phi+|\eta_s\rangle\cos\phi.
\end{eqnarray}
For the calculations we use the experimental value of the mixing angle
$\phi=(41.4\pm0.3\pm0.7\pm0.6)^\circ$ \cite{KLOE} and
neglect the possible glue content in these mesons. The wave functions of $\eta_q$
and $\eta_s$ mesons are known in our model from previous light meson mass
spectra investigations \cite{lm}. The calculated masses of the pure $\eta_q$ and
$\eta_s$ are $M_{\eta_q}=154$~MeV and $M_{\eta_s}=743$~MeV \cite{lm}, while the
masses of the mixed states $\eta$ and $\eta'$ are close to the
experimental values.

\begin{figure}
\centering
  \includegraphics[width=8cm]{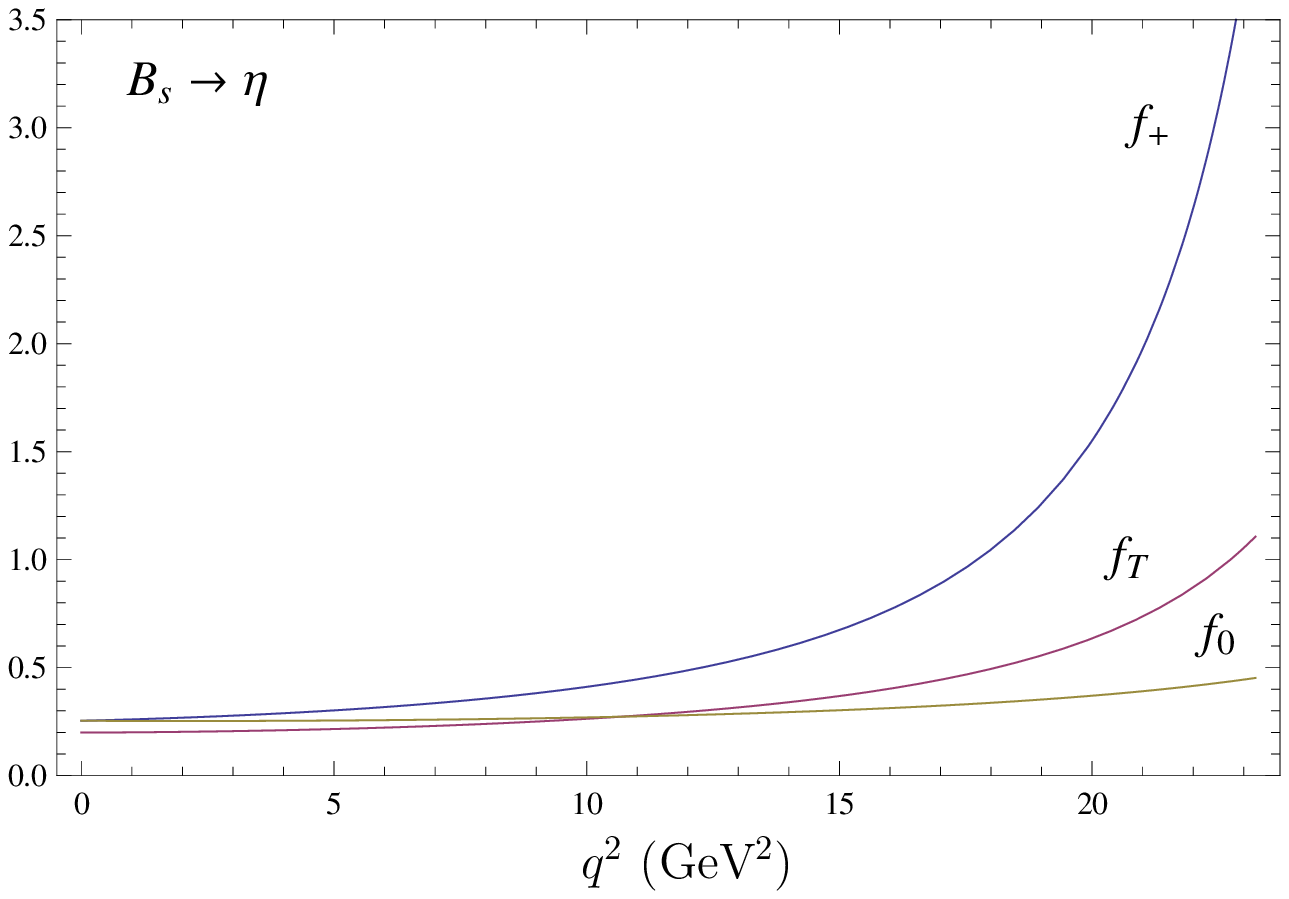}\ \ \ \includegraphics[width=8cm]{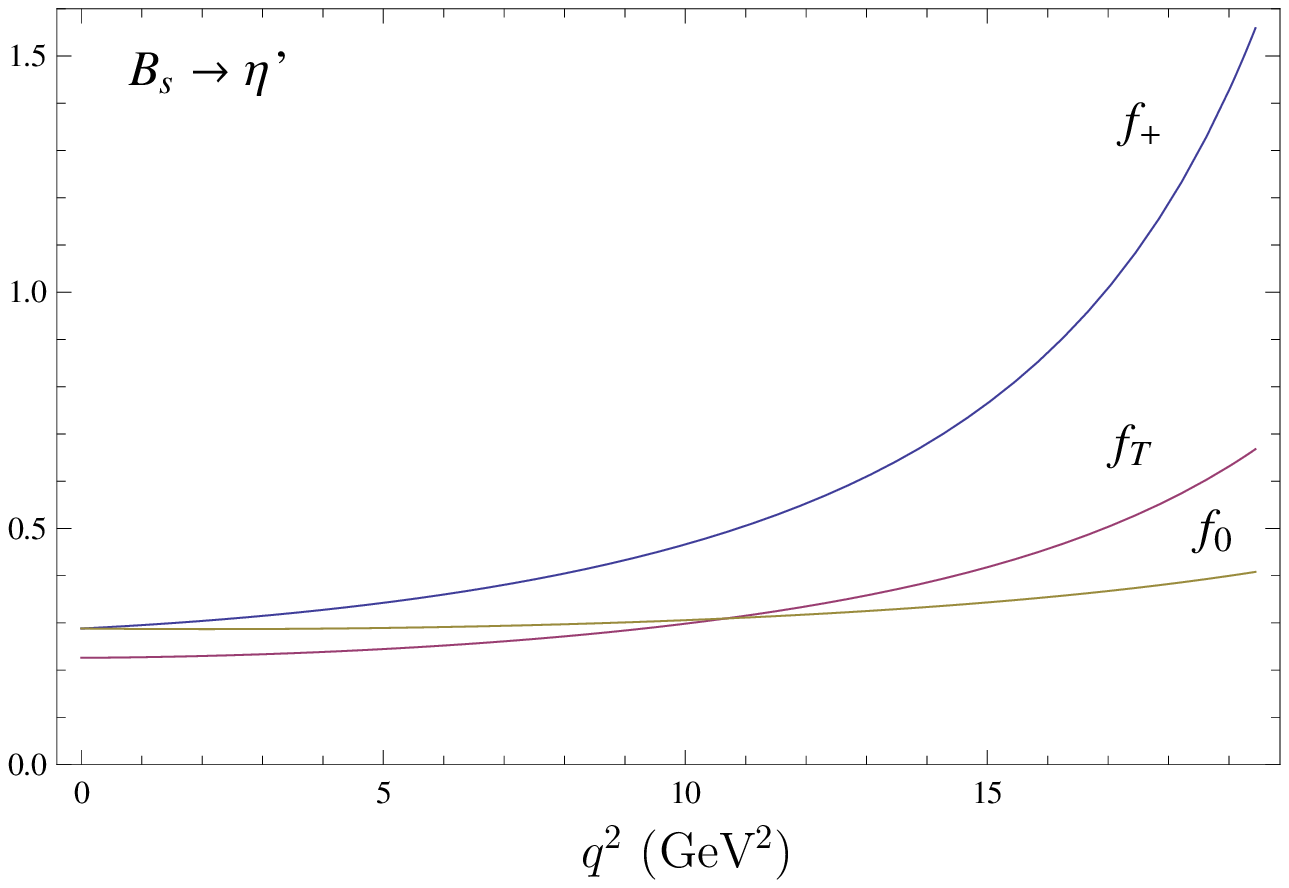}
\caption{Form factors of the weak $B_s\to \eta$ and $B_s\to \eta'$ transitions.    } 
\label{fig:ffk}
\end{figure}

\begin{figure}
\centering
  \includegraphics[width=8cm]{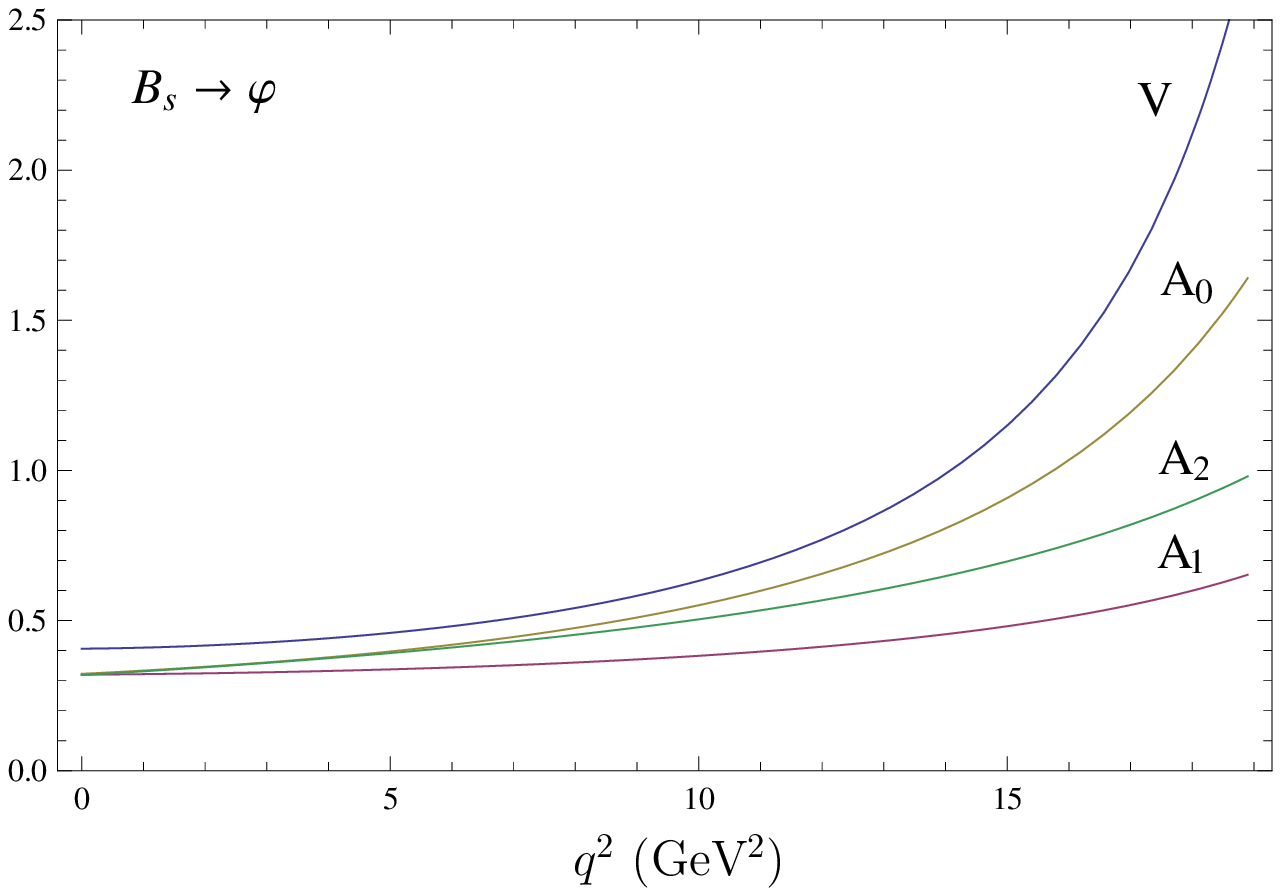}\ \
 \ \includegraphics[width=8cm]{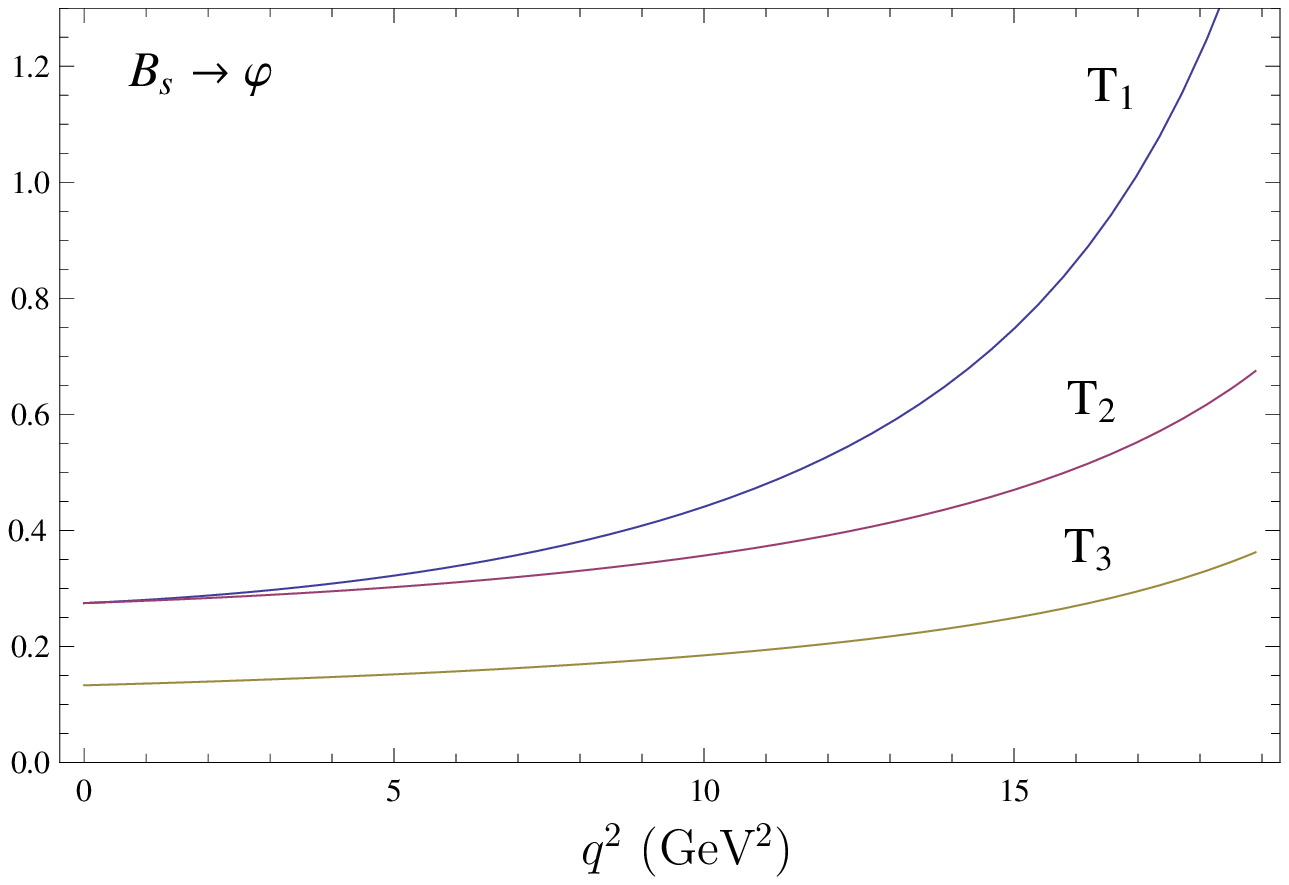}\\
\caption{Form factors of the weak $B_s\to \varphi$ transitions.    } 
\label{fig:ffks}
\end{figure}

Now we compare the form factor decompositions (\ref{eq:pff1})--(\ref{eq:vff4}) with the results of the calculations of
the weak current matrix element in our model, based on the methods described in the
previous section. This allows us to explicitly determine the form
factors in the whole accessible kinematical range through the overlap
integrals of the meson wave functions. The corresponding expressions can be
found in Refs.~\cite{bcdecay,brare}. For the numerical evaluations of
the overlap integrals we use the quasipotential wave
functions of $B_s$ and $\eta(\varphi)$ mesons previously obtained in
their mass spectra studies \cite{hlm,lm}. The calculated form factors are
plotted in Figs.~\ref{fig:ffk}  and \ref{fig:ffks}.

For the comparison of the obtained form factors with experiment and
other theoretical calculations it is important to have approximate analytic
expressions for them. Our analysis shows that the weak $B_s\to\eta(\varphi)$ transition form factors can be well fitted  by the following formulas
\cite{ms,bdecays}: 

(a) $F(q^2)= \{f_+(q^2),f_T(q^2),V(q^2),A_0(q^2),T_1(q^2)\}$ 
\begin{equation}
  \label{fitfv}
  F(q^2)=\frac{F(0)}{\displaystyle\left(1-\frac{q^2}{M^2}\right)
    \left(1-\sigma_1 
      \frac{q^2}{M_{B^*_s}^2}+ \sigma_2\frac{q^4}{M_{B^*_s}^4}\right)},
\end{equation}

(b) $F(q^2)=\{f_0(q^2), A_1(q^2),A_2(q^2),T_2(q^2),T_3(q^2)\}$
\begin{equation}
  \label{fita12}
  F(q^2)=\frac{F(0)}{\displaystyle \left(1-\sigma_1
      \frac{q^2}{M_{B^*_s}^2}+ \sigma_2\frac{q^4}{M_{B^*_s}^4}\right)},
\end{equation}
where $M=M_{B^*_s}$ for the form factors $f_+(q^2),f_T(q^2),V(q^2),T_1(q^2)$ and
$M=M_{B_s}$ for the form factor $A_0(q^2)$. The obtained values of $F(0)$ and
$\sigma_{1,2}$ are given in Table~\ref{hff}. The quality of such approximation is
rather high, the deviation from the
calculated form factors does not exceed 1\%. The rough
estimate of the total uncertainty of the form factors within
our model gives its values of order of 5\%.  The subleading
contributions (\ref{gamma2}) to the decay matrix elements in the region of small
recoils are the main source of these uncertainties. 

\begin{table}
\caption{Calculated form factors of weak $B_s\to \eta_s$ and $B_s\to \varphi$ transitions. Form factors $f_+(q^2)$, $f_T(q^2)$ $V(q^2)$,
  $A_0(q^2)$, $T_1(q^2)$ are fitted by Eq.~(\ref{fitfv}), and form factors $f_0(q^2)$,
  $A_1(q^2)$, $A_2(q^2)$, $T_2(q^2)$, $T_3(q^2)$  are fitted by Eq.~(\ref{fita12}).  }
\label{hff}
\begin{ruledtabular}
\begin{tabular}{ccccccccccc}
   &\multicolumn{3}{c}{{$B_s\to \eta_s$}}&\multicolumn{7}{c}{{\  $B_s\to \varphi$\
     }}\\
\cline{2-4} \cline{5-11}
& $f_+$ & $f_0$& $f_T$& $V$ & $A_0$ &$A_1$&$A_2$& $T_1$ & $T_2$& $T_3$\\
\hline
$F(0)$          &0.384 &0.384 &  0.301 & 0.406 & 0.322& 0.320 & 0.318 & 0.275& 0.275& 0.133\\
$F(q^2_{\rm max})$&3.31  &0.604 &  1.18 & 2.74& 1.64& 0.652 &  0.980 & 1.47& 0.675& 0.362\\
$\sigma_1$      &$-0.347$&$-0.120$& $-0.897$& $-0.861$ &$-0.104$&  0.133&
1.11  &$-0.491$&$0.396$& $0.639$\\
$\sigma_2$      &$-1.55$&$-0.849$&$-1.34$&$-2.74$&$-1.19$&$-1.02$&
0.105  &$-1.90$&$-0.811$& $-0.531$\\
\end{tabular}
\end{ruledtabular}
\end{table}

In Table~\ref{compbsff} we confront our predictions for the form
factors of rare weak $B_s\to  \eta_s$ and $B_s\to  \varphi$
transitions at maximum recoil  ($q^2=0$) with previous calculations
\cite{bz,akllsww,ikkss,ms,llw,lww,wzz,gl,swyz} within various
theoretical approaches. The different versions of light-cone sum rules
are used in Refs.~\cite{bz,wzz}. The authors of Refs.~\cite{akllsww,llw}
employ the perturbative QCD approach, while the covariant constituent
quark model with the infrared confinement is used in
Ref.~\cite{ikkss}. Considerations in Ref.~\cite{ms} are based on the
constituent quark model and relativistic dispersion approach. The
light-cone quark model calculations are performed in
Refs.~\cite{lww,gl} and the six-quark effective Hamiltonian model
is employed in Ref.~\cite{swyz}. Comparison of the results presented
in this table shows that, although there are some 
differences between predictions, in general there is a reasonable
agreement between the values of these form factors at zero recoil
calculated using significantly different theoretical methods. However,
most of the discussed  approaches allow the form factor calculation at
the single point only or in some limited region of the recoil momentum,
then some model extrapolation to the whole kinematical range should be
used. The important advantage of our model is the explicit
determination of the momentum dependence of the form factors without
any additional assumptions. 

\begin{figure}
\centering
  \includegraphics[width=8cm]{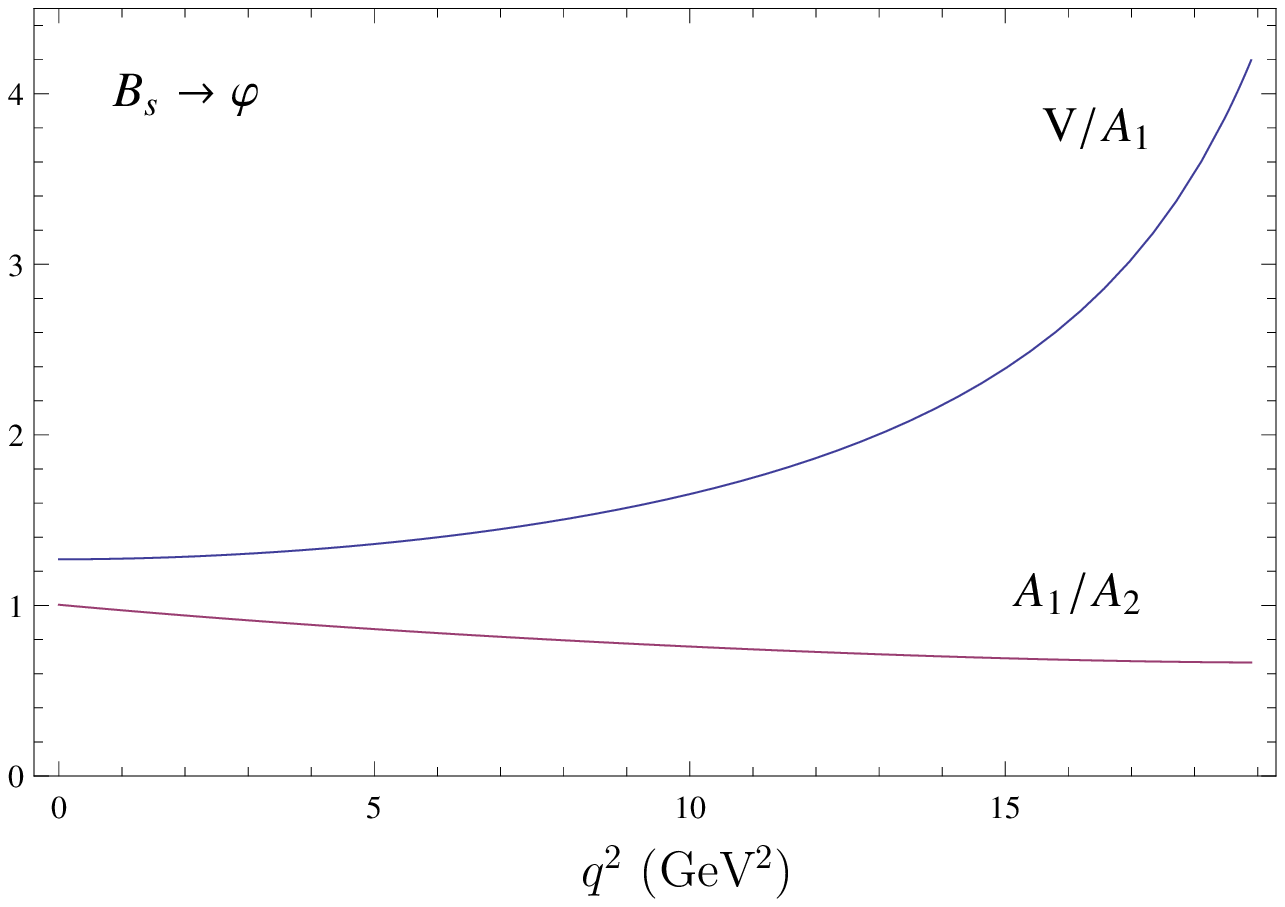}
\caption{The form factor ratios for the weak $B_s\to \varphi$ transitions.    } 
\label{fig:ratio}
\end{figure}

Recently in Ref.~\cite{hh} the method of extracting the $B\to
K^*$  transition form factors from available experimental data was proposed. It
allows one to extract the ratios of the form factors $V(q^2)/A_1(q^2)$
and $A_1(q^2)/A_2(q^2)$ from the experimental data on angular
distributions in this decay. Similar approach can be, in principle, applied for the
 $B_s\to \varphi$  transition. In Fig.~\ref{fig:ratio} we give our
predictions for the corresponding form factor ratios. Using $SU(3)$
symmetry arguments one can expect that the ratios of these form
factors should have similar $q^2$ behaviour for $B$ and $B_s$
decays. Indeed, we observe the 
qualitative agreement of these ratios with the ones found in Ref.~\cite{hh}.

In the following sections we apply the calculated form factors for the
consideration of the rare semileptonic, radiative and nonleptonic
decays of $B_s$ mesons.

\begin{sidewaystable}
\caption{Comparison of theoretical predictions for the form factors of 
  weak $B_s\to  \eta_s$ and $B_s\to  \varphi$ transitions at maximum
  recoil point $q^2=0$.  }
\label{compbsff}
\begin{ruledtabular}
\begin{tabular}{ccccccccc}
     & $f_+(0)$& $f_T(0)$ & $V(0)$ & $A_0(0)$ &$A_1(0)$&$A_2(0)$ &$T_1(0)$&$T_3(0)$ \\
\hline
This paper   &$0.384\pm0.019$  & $0.301\pm0.015$  &$0.406\pm0.020$  &$0.322\pm0.016$ &$0.320\pm0.016$  &$0.318\pm0.016$ & $0.275\pm0.014$& $0.133\pm0.006$\\
\cite{bz}   &  &   &$0.434\pm0.035$ & $0.474\pm0.033$ & $0.311\pm0.030$& $0.234\pm0.028$& $0.349\pm0.033$& $0.175\pm0.018$\\
\cite{akllsww} &$0.36\pm0.07$ & &$0.25\pm0.05$&$0.30\pm0.06$ &$0.19\pm0.04$
 & \\
\cite{ikkss} & & &0.32 & & 0.29 & 0.28 & 0.28\\
\cite{ms}& 0.36&0.36 &$0.44$ &0.42 & 0.34 &0.31&0.38&0.26
  \\
\cite{llw} & & &$0.26\pm0.07$&$0.31^{+0.08}_{-0.07}$& $0.18^{+0.06}_{-0.05}$& $0.12\pm0.03$& $0.23^{+0.06}_{-0.05}$& $0.19^{+0.06}_{-0.05}$\\
\cite{lww} &0.288 & &0.329 &0.279&0.232 &0.210&0.276&0.170\\
\cite{wzz} &$0.281\pm0.015$ &$0.282\pm0.016$ &$0.339\pm0.017$
&$0.269\pm0.014$&$0.271\pm0.014$&$0.212\pm0.011$&$0.299\pm0.016$&$0.191\pm0.010$
\\
\cite{gl} &0.357 & 0.365 &0.445& &0.343&0.310&0.380\\
\cite{swyz} && 
&$0.259^{+0.082}_{-0.037}$&$0.311^{+0.098}_{-0.049}$&$0.194^{+0.054}_{-0.029}$
\end{tabular}
\end{ruledtabular}
\end{sidewaystable}

\section{Rare semileptonic $B_s\to \eta(\varphi)l^+l^-$ decays}
\label{sec:rard}

First we consider the rare semileptonic decays. In the
following calculations the usual factorization of short-distance (described by
Wilson coefficients) and long-distance (which matrix elements are proportional to hadronic form factors)
 contributions in the effective Hamiltonian for the $b\to s$
transitions is employed \cite{bhi}       
\begin{equation}
  \label{eq:heff}
  {\cal H}_{\rm eff} =-\frac{4G_F}{\sqrt{2}}V_{ts}^*V_{tb}\sum_{i=1}^{10}c_i{\cal
      O}_i,
\end{equation}
where $G_F$ is the Fermi constant, $V_{tj}$ are
Cabibbo-Kobayashi-Maskawa matrix elements, $c_i$ are the Wilson coefficients
and ${\cal O}_i$ are the standard model  operators.

Then the matrix element of the $b\to s l^+l^-$ transition amplitude between meson states can be written \cite{fgikl,abhh} in
the following form
\begin{eqnarray}
  \label{eq:mtl}
  {\cal M}(B_s\to \eta l^+l^-)&=&\frac{G_F\alpha}{2\sqrt{2}\pi}
  |V_{ts}^*V_{tb}|\left[T^{(1)}_\mu(\bar l\gamma^\mu l)+T^{(2)}_\mu
    (\bar l\gamma^\mu \gamma_5l)\right],\cr
 {\cal M}(B_s\to \varphi l^+l^-)&=&\frac{G_F\alpha}{2\sqrt{2}\pi}
  |V_{ts}^*V_{tb}|\left[\epsilon^{\dag\nu}T^{(1)}_{\mu\nu}(\bar l\gamma^\mu l)+\epsilon^{\dag\nu}T^{(2)}_{\mu\nu}
    (\bar l\gamma^\mu \gamma_5l)\right],
\end{eqnarray}
where $T^{(i)}$ are expressed through the form factors and the Wilson
coefficients. These amplitudes can be written in the helicity
basis $\varepsilon^{\mu}(m)$ as follows (see \cite{fgikl})

(a) $B\to \eta^{(')}$ transition:
\begin{equation}
  \label{eq:hap}
  H^{(i)}_m=\varepsilon^{\dag \mu}(m)T^{(i)}_\mu,
\end{equation}
where  
\begin{eqnarray}
  \label{eq:hp}
  H^{(i)}_\pm&=&0,\cr
H^{(1)}_0&=&\frac{\lambda^{1/2}}{\sqrt{q^2}}\left[c_9^{eff} f_+(q^2)+ c_7^{eff}\frac{2m_b}{M_{B_s}+M_\eta}f_T(q^2)\right],\cr
H^{(2)}_0&=&\frac{\lambda^{1/2}}{\sqrt{q^2}} c_{10} f_+(q^2),\cr
H^{(1)}_t&=&\frac{M_{B_s}^2-M_\eta^2}{\sqrt{q^2}}c_9^{eff} f_0(q^2),\cr
H^{(2)}_t&=&\frac{M_{B_s}^2-M_\eta^2}{\sqrt{q^2}}c_{10} f_0(q^2).
\end{eqnarray}
Here $\lambda\equiv
\lambda(M_{B_s}^2,M_F^2,q^2)=M_{B_s}^4+M_F^4+q^4-2(M_{B_s}^2M_F^2+M_F^2q^2+M_{B_s}^2q^2)$
and the subscripts $\pm,0,t$ denote transverse, longitudinal and time helicity
components, respectively.

(b) $B\to \varphi$ transition:
\begin{equation}
  \label{eq:hav}
  H^{(i)}_m=\varepsilon^{\dag \mu}(m)\epsilon^{\dag\nu}T^{(i)}_{\mu\nu},
\end{equation}
where  $\epsilon^{\nu}$ is the polarization vector of the vector $\varphi$ meson and
\begin{eqnarray}
  \label{eq:hpv}
  H^{(1)}_\pm&=&-(M_{B_s}^2-M_\varphi^2)\left[c_9^{eff}\frac{A_1(q^2)}{M_{B_s}-M_\varphi}+
    \frac{2m_b}{q^2}c_7^{eff}T_2(q^2)\right]\cr
&&\pm \lambda^{1/2}
\left[c_9^{eff}\frac{V(q^2)}{M_{B_s}+M_\varphi}+
    \frac{2m_b}{q^2}c_7^{eff}T_1(q^2)\right],\cr
 H^{(2)}_\pm&=&c_{10}\left[-(M_{B_s}+M_\varphi)A_1(q^2)\pm \lambda^{1/2}
\frac{V(q^2)}{M_B+M_\varphi}\right],\cr
H^{(1)}_0&=&-\frac{1}{2M_\varphi\sqrt{q^2}}\Biggl[c_9^{eff}\left\{(M_{B_s}^2-M_\varphi^2-q^2)
    (M_{B_s}+M_\varphi)A_1(q^2)-\frac{\lambda}{M_{B_s}+M_\varphi}A_2(q^2)\right\}\cr
&&+2m_b c_7^{eff}\left\{(M_{B_s}^2+3M_\varphi^2-q^2)
 T_2(q^2)-\frac{\lambda}{M_{B_s}^2-M_\varphi^2}T_3(q^2)\right\}\Biggr],\cr
H^{(2)}_0&=&-\frac{1}{2M_V\sqrt{q^2}}c_{10}\left[(M_{B_s}^2-M_\varphi^2-q^2)
    (M_{B_s}+M_\varphi)A_1(q^2)-\frac{\lambda}{M_{B_s}+M_\varphi}A_2(q^2)\right],\cr
H^{(1)}_t&=&-\frac{\lambda^{1/2}}{\sqrt{q^2}}c_9^{eff} A_0(q^2),\cr
H^{(2)}_t&=&-\frac{\lambda^{1/2}}{\sqrt{q^2}}c_{10} A_0(q^2).
\end{eqnarray}
The values of the Wilson coefficients $c_i$ and of the effective
Wilson coefficient $c_7^{eff}$  are taken from Ref.~\cite{wc}. The 
effective  Wilson coefficient $ c_9^{\rm eff}$ contains additional
pertubative and long-distance contributions   
\begin{equation}
 \label{eq:ceff9}
  c_9^{\rm eff}=c_9+{\cal Y}_{\rm pert}(q^2)+{\cal Y}_{\rm BW}(q^2).
\end{equation}
The perturbative part is equal to
\begin{eqnarray}
  \label{eq:ypert}
{\cal Y}_{\rm pert}(q^2)&=&h\left(\frac{m_c}{m_b},\frac{q^2}{m_b^2}\right)(3c_1+c_2+3c_3+c_4+3c_5+c_6)\cr
&&-
\frac12 h\left(1,\frac{q^2}{m_b^2}\right)(4c_3+4c_4+3c_5+c_6)\cr
&&-\frac12
h\left(0,\frac{q^2}{m_b^2}\right)(c_3+3c_4)+\frac29(3c_3+c_4+3c_5+c_6),
\end{eqnarray}
where
\begin{eqnarray*} 
h(\frac{m_c}{m_b},  \frac{q^2}{m_b}) & = & 
- \frac{8}{9}\ln\frac{m_c}{m_b} +
\frac{8}{27} + \frac{4}{9} x 
-  \frac{2}{9} (2+x) |1-x|^{1/2} \left\{
\begin{array}{ll}
 \ln\left| \frac{\sqrt{1-x} + 1}{\sqrt{1-x} - 1}\right| - i\pi, &
 x \equiv \frac{4 m_c^2}{ q^2} < 1,  \\
 & \\
2 \arctan \frac{1}{\sqrt{x-1}}, & x \equiv \frac
{4 m_c^2}{ q^2} > 1,
\end{array}
\right. \\
h(0, \frac{q^2}{m_b} ) & = & \frac{8}{27} - 
\frac{4}{9} \ln\frac{q^2}{m_b} + \frac{4}{9} i\pi.
\end{eqnarray*}
The long-distance (nonperturbative) contributions are assumed to originate from the $c\bar c$ resonances ($J/\psi, \psi'\dots$) and  have
a usual Breit-Wigner structure:
\begin{equation}
  \label{eq:ybw}
{\cal Y}_{\rm BW}(q^2)=\frac{3\pi}{\alpha^2} \sum_{V_i=J/\psi,\psi(2S)\dots}\frac{\Gamma(V_i\to l^+l^-)M_{V_i} }{M_{V_i}^2-q^2-iM_{V_i}\Gamma_{V_i}}.
\end{equation}
We include contributions of the vector $V_i(1^{--})$ charmonium states: $J/\psi$,
$\psi(2S)$, $\psi(3770)$, $\psi(4040)$, $\psi(4160)$ and $\psi(4415)$,
with their masses ($M_{V_i}$), leptonic [$\Gamma(V_i\to l^+l^-)$] and
total ($\Gamma_{V_i}$) decay widths taken from PDG \cite{pdg}.

The differential decay rate can be written in terms of the helicity
amplitudes \cite{fgikl} as follows
\begin{eqnarray}
  \label{eq:dgamma}
  \frac{d\Gamma(B_s\to \eta(\varphi)l^+l^-)}{dq^2}&=&\frac{G_F^2}{(2\pi)^3}
  \left(\frac{\alpha |V_{ts}^*V_{tb}|}{2\pi}\right)^2
  \frac{\lambda^{1/2}q^2}{48M_{B_s}^3} \sqrt{1-\frac{4m_l^2}{q^2}}
  \Biggl[H^{(1)}H^{\dag(1)}\left(1+\frac{2m_l^2}{q^2}\right)\cr
&& +
    H^{(2)}H^{\dag(2)}\left(1-\frac{4m_l^2}{q^2}\right) +\frac{2m_l^2}{q^2}3 H^{(2)}_tH^{\dag(2)}_t\Biggr],
\end{eqnarray}
where $m_l$ is the lepton mass and
\begin{equation}
  \label{eq:hh}
  H^{(i)}H^{\dag(i)}\equiv H^{(i)}_+H^{\dag(i)}_++H^{(i)}_-H^{\dag(i)}_-+H^{(i)}_0H^{\dag(i)}_0.
\end{equation}

The other convenient observables for the $B_s\to
\varphi\mu^+\mu^-$ ($\varphi\to K^+K^-$) decay,  used in
measurements,  are the forward-backward asymmetry $A_{FB}$ and the
longitudinal fraction of the polarization of the vector $\varphi$
meson $ F_L$. They enter the differential decay distributions in $\cos
\theta_K$
\begin{equation}
  \label{eq:dgth}
  \frac1{\Gamma}\frac{d\Gamma(B_s\to \varphi\mu^+\mu^-)}{d\cos \theta_K}=
  \frac32 F_L\cos^2 \theta_K+\frac34 (1- F_L)(1-\cos^2
  \theta_K), 
\end{equation}
and in $\cos \theta_\mu$
\begin{equation}
  \label{eq:dgtheta}
  \frac1{\Gamma}\frac{d\Gamma(B_s\to \varphi\mu^+\mu^-)}{d\cos \theta_\mu}=
  \frac34 F_L(1-\cos^2 \theta_\mu)+\frac38 (1-F_L)(1+\cos^2
  \theta_\mu) +A_{FB}\cos \theta_\mu, 
\end{equation}
where $\theta_K$ is the angle between the $K^+$ direction and the
direction opposite to the $B_s$ meson in the $\varphi$ rest frame, and 
$\theta_\mu$ is the angle between the $\mu^+$ and the opposite of the
$B$ direction in the dilepton rest frame.

These observables are expressed through the helicity amplitudes in the
following way.

(a) The forward-backward asymmetry
\begin{equation}
  \label{eq:afb}
  A_{FB}=\frac34 \sqrt{1-\frac{4m_l^2}{q^2}}\frac{ {\rm Re}(H^{(1)}_+H^{\dag(2)}_+)-{\rm Re}(H^{(1)}_-H^{\dag(2)}_-)}{H^{(1)}H^{\dag(1)}\left(1+\frac{2m_l^2}{q^2}\right) +
    H^{(2)}H^{\dag(2)}\left(1-\frac{4m_l^2}{q^2}\right) +\frac{2m_l^2}{q^2}3 H^{(2)}_tH^{\dag(2)}_t}.
\end{equation}

(b) The longitudinal polarization fraction of the vector $\varphi$ meson
\begin{equation}
  \label{eq:fl}
  F_L=\frac{H^{(1)}_0H^{\dag(1)}_0\left(1+\frac{2m_l^2}{q^2}\right) +
    H^{(2)}_0H^{\dag(2)}_0\left(1-\frac{4m_l^2}{q^2}\right) +\frac{2m_l^2}{q^2}3 H^{(2)}_tH^{\dag(2)}_t }{H^{(1)}H^{\dag(1)}\left(1+\frac{2m_l^2}{q^2}\right) +
    H^{(2)}H^{\dag(2)}\left(1-\frac{4m_l^2}{q^2}\right) +\frac{2m_l^2}{q^2}3 H^{(2)}_tH^{\dag(2)}_t}.
\end{equation}
They are the most popular quantities for the rare weak decays, since they  can be
determined experimentally using the angular analysis.

\begin{figure}
  \centering
 \includegraphics[width=7.8cm]{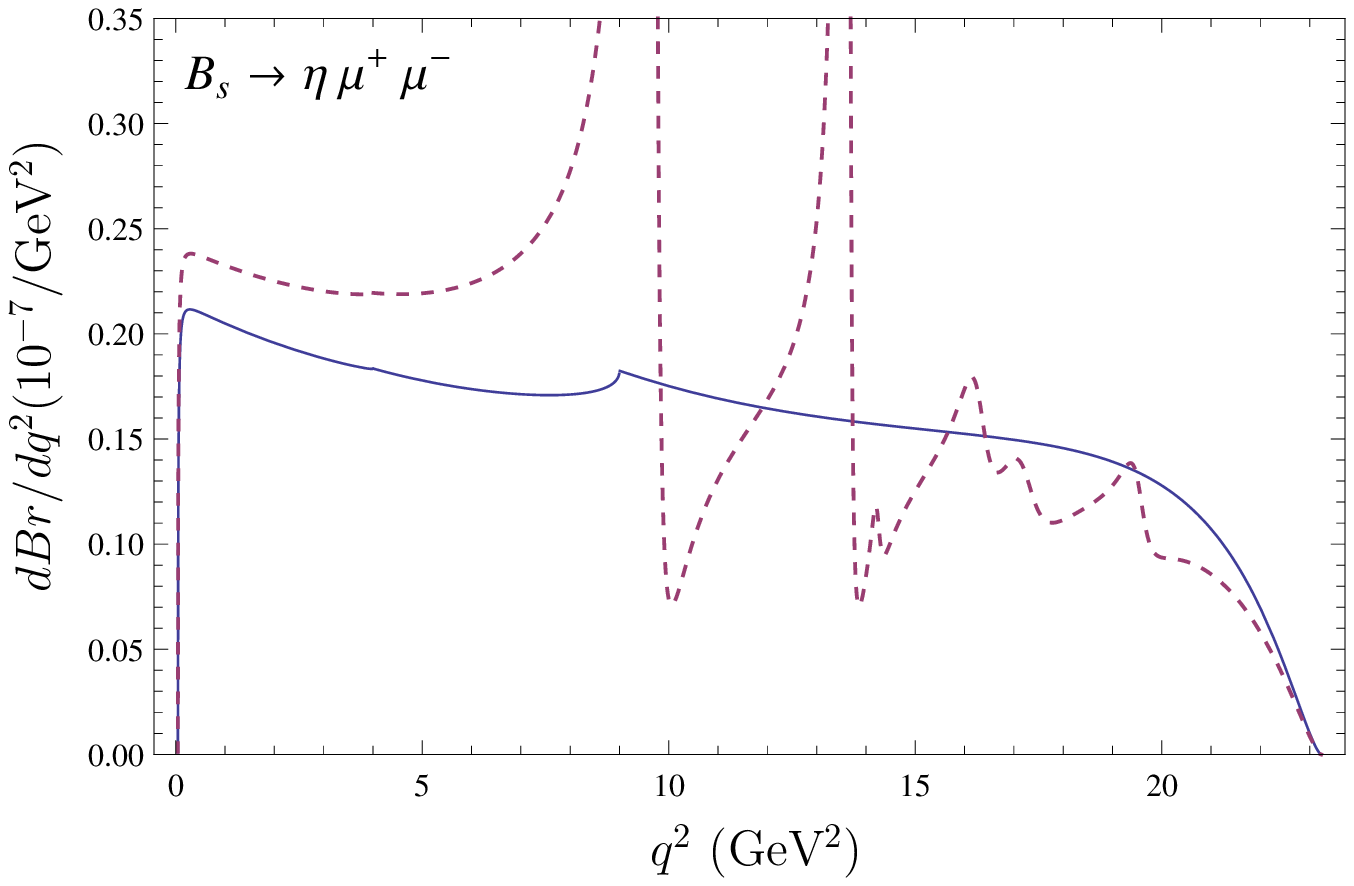} \ \ \   \includegraphics[width=7.8cm]{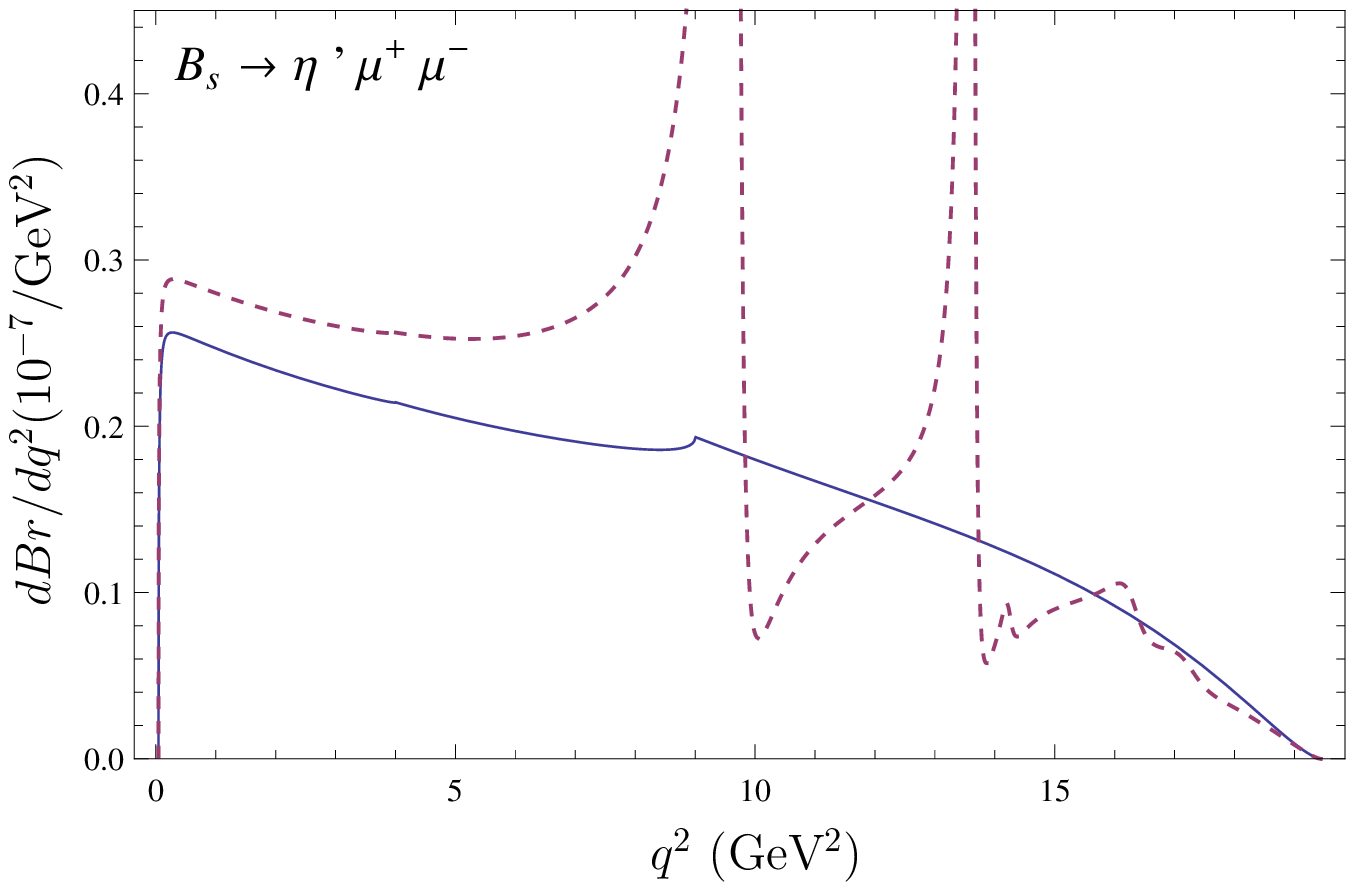}\\
 \includegraphics[width=8cm]{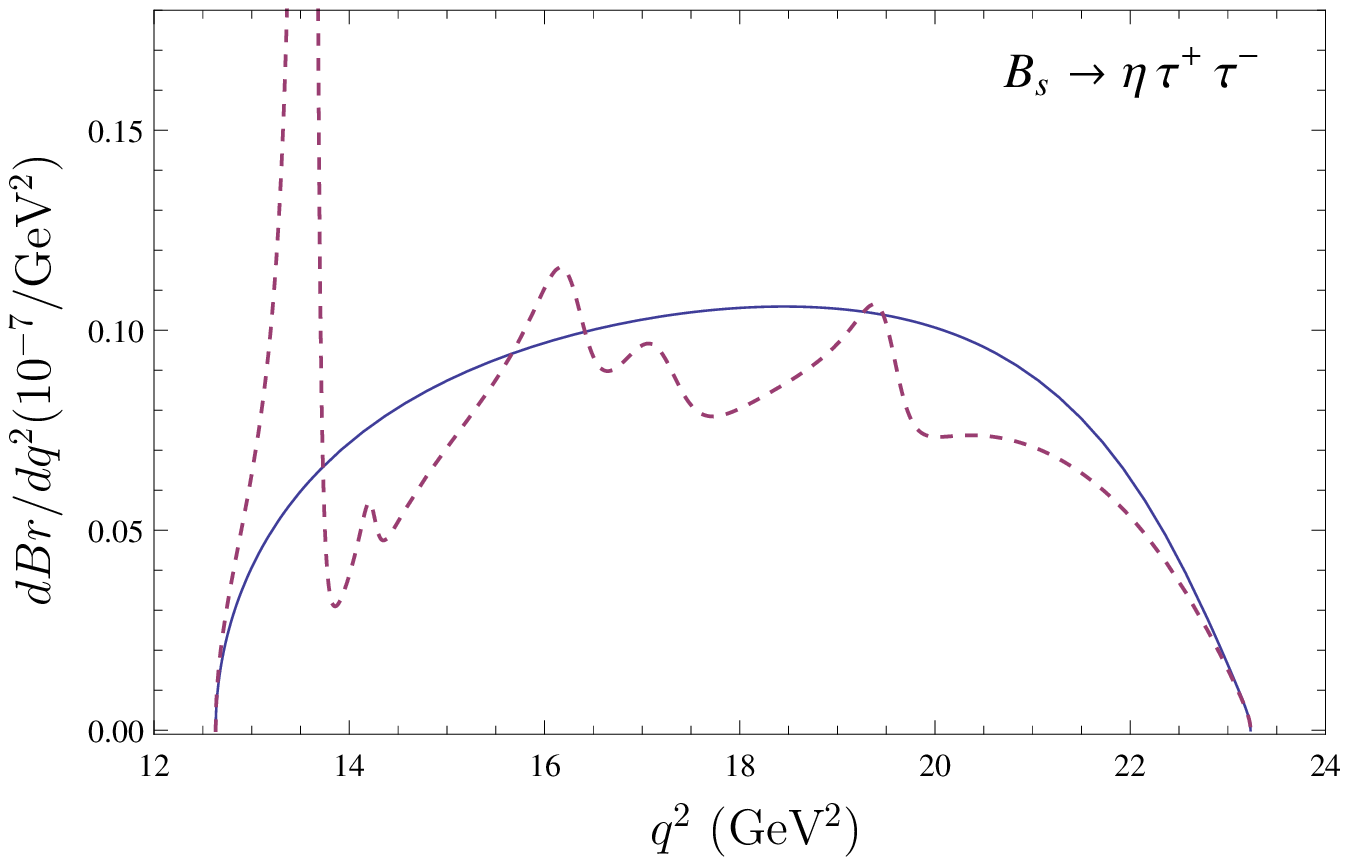}\ \ \  \includegraphics[width=8cm]{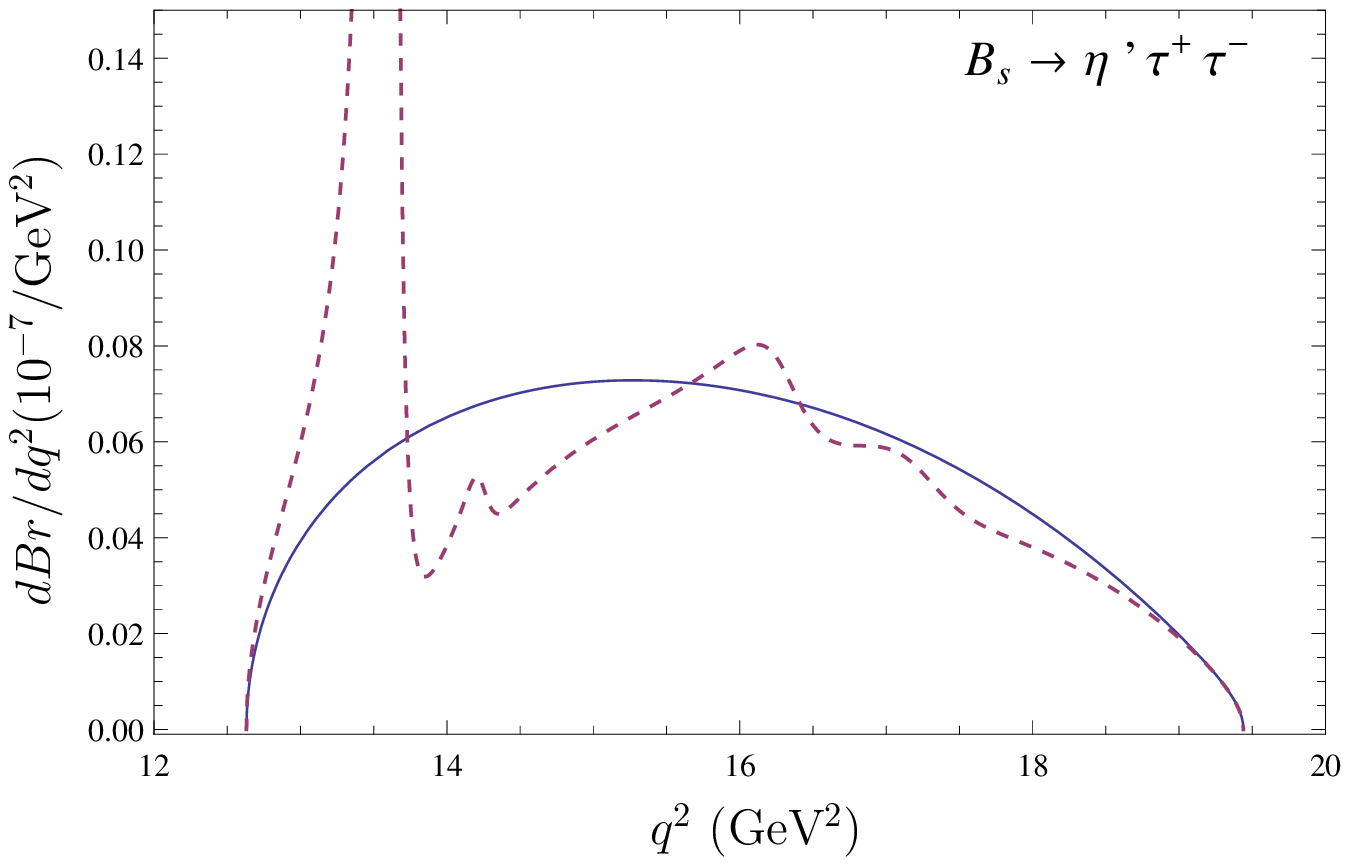}
  \caption{Theoretical predictions for the differential
    branching fractions $d Br(B_s \to \eta^{(')}l^+l^-)/d q^2$. Nonresonant
    and resonant results are plotted by solid and dashed lines,
    respectively. }
  \label{fig:brbseta}
\end{figure}

\begin{figure}
  \centering
 \includegraphics[width=7.8cm]{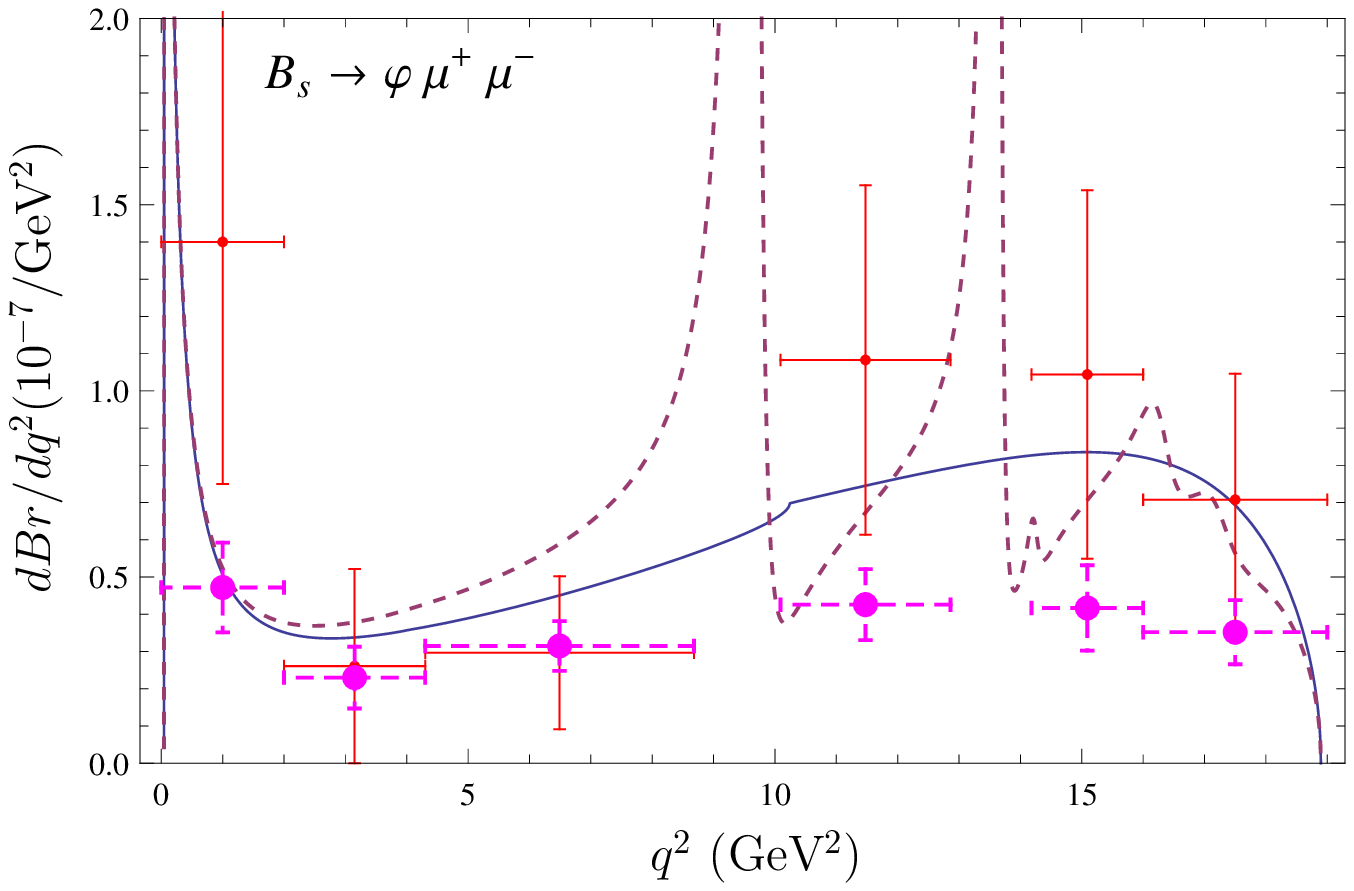} \ \ \  \includegraphics[width=7.8cm]{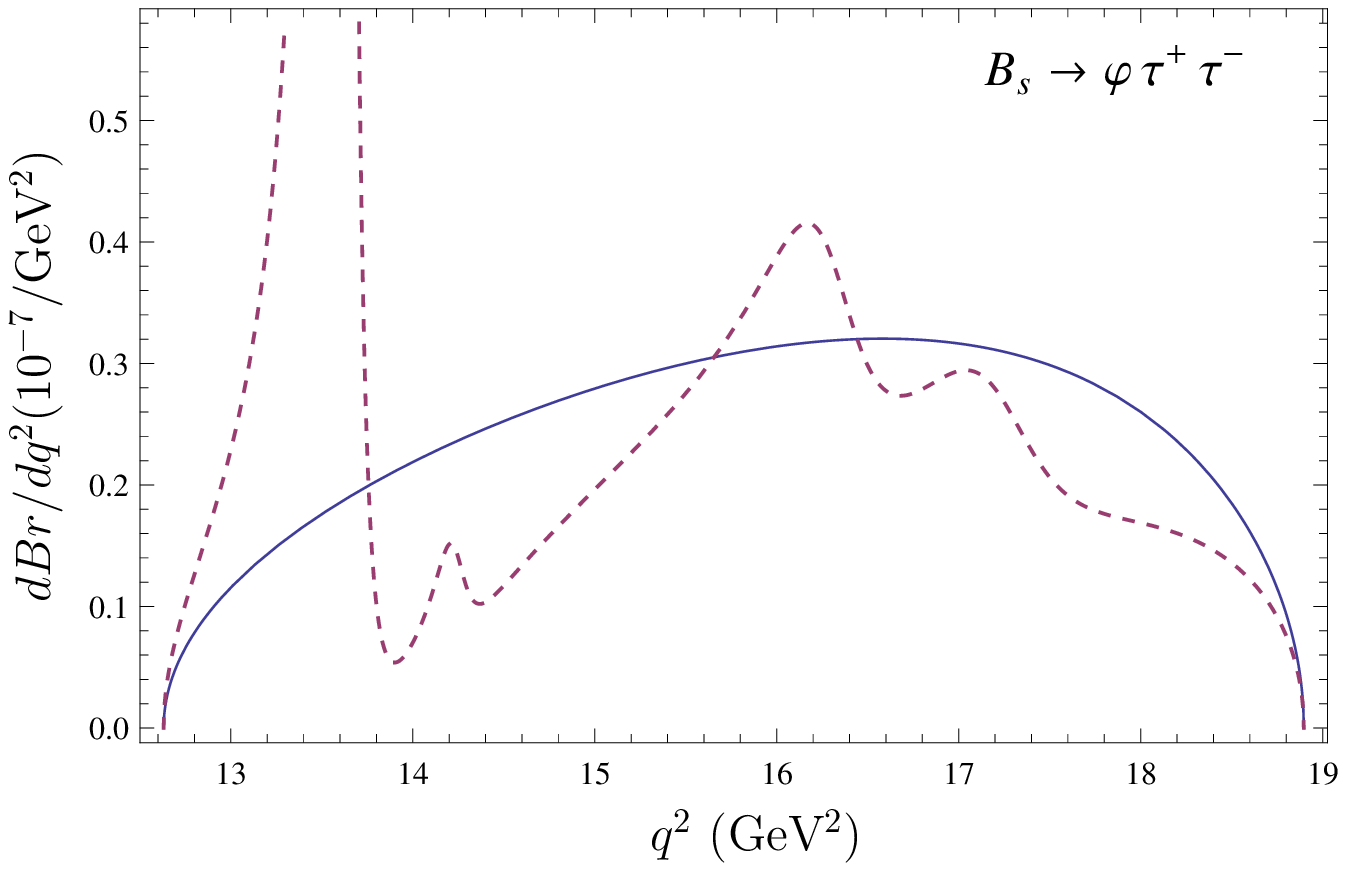}
\caption{Comparison of theoretical predictions for the differential
    branching fractions $d Br(B_s \to \varphi \mu^+\mu^-)/d q^2$  with available experimental data. Nonresonant
    and resonant results are plotted by solid and dashed lines,
    respectively. CDF data are
    given by dots with solid error bars, while LHCb data are presented by
    filled circles with dashed error bars.}
  \label{fig:brbsphi}
\end{figure}

\begin{figure}
  \centering
\includegraphics[width=7.8cm]{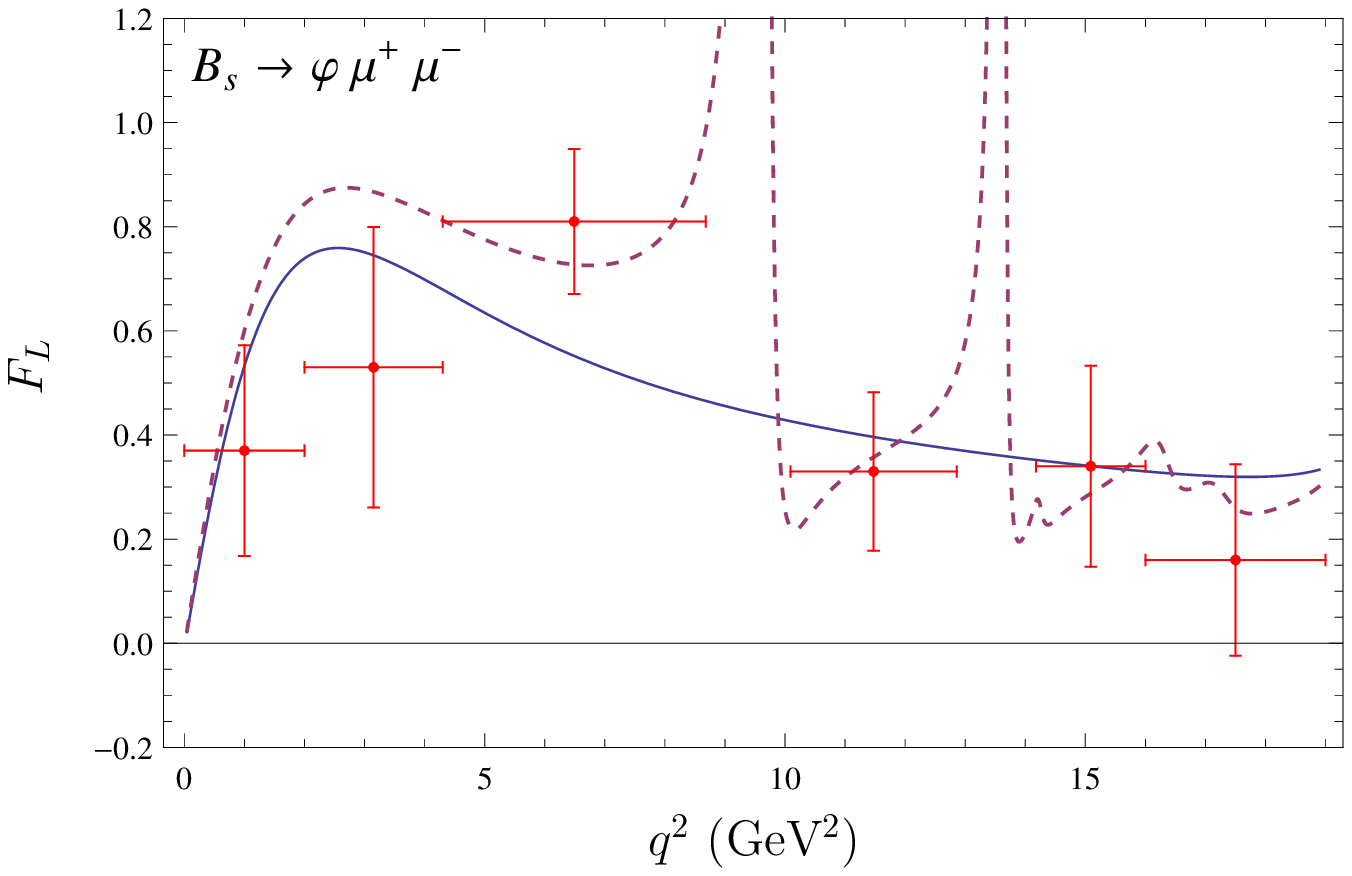}\ \ \ \
\  \includegraphics[width=7.8cm]{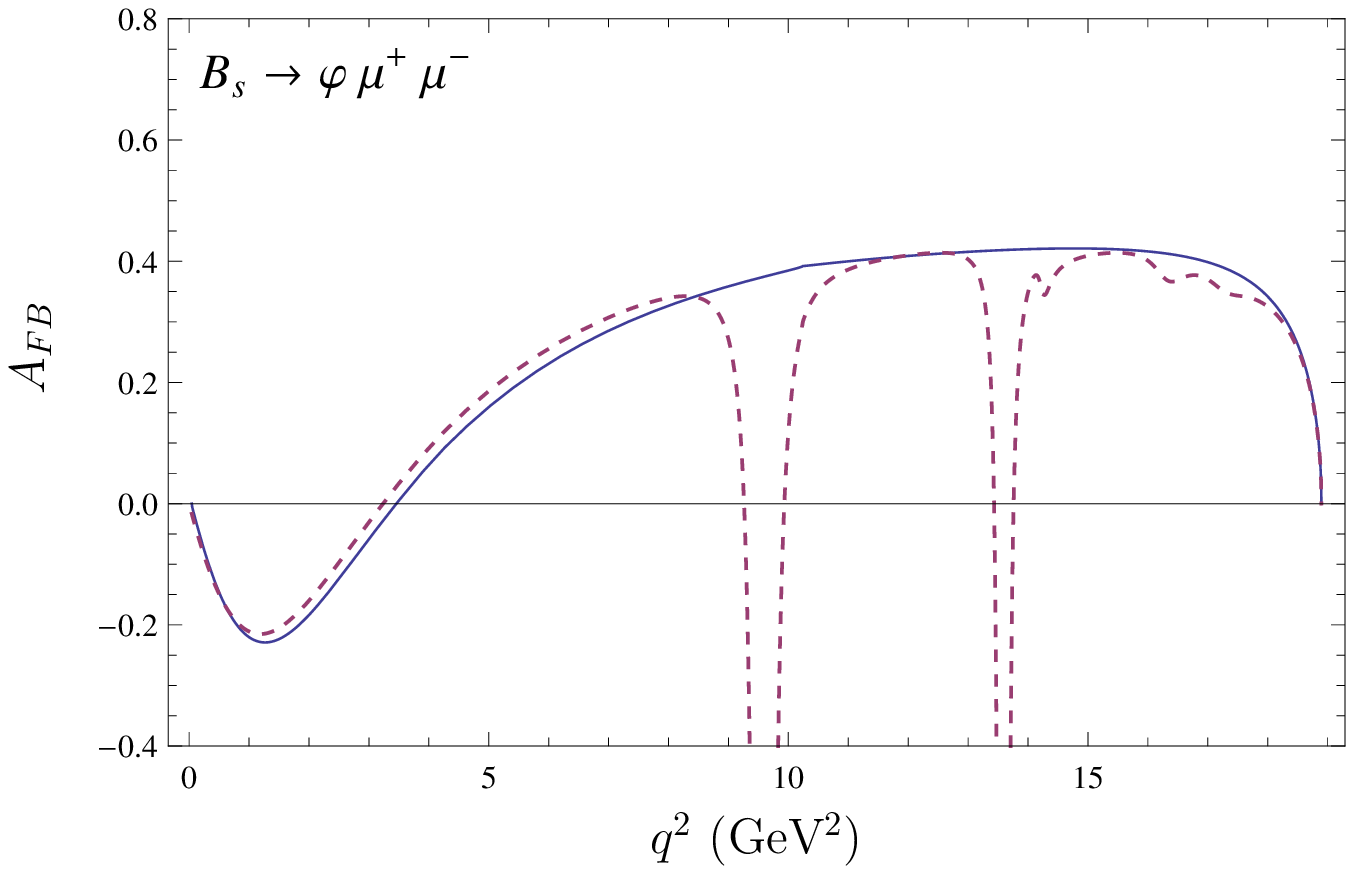}

  \caption{Comparison of theoretical predictions for the $\varphi$ longitudinal polarization
    $F_L$ and muon forward-backward asymmetry $A_{FB}$ for the rare $B_s \to
    \varphi \mu^+\mu^-$ decays with available experimental data. Nonresonant
    and resonant results are plotted by solid and dashed lines,
    respectively. LHCb data are
    given by dots with solid error bars.}
  \label{fig:brbk}
\end{figure}

Now we substitute into these expressions the rare $B_s$ decay form
factors calculated in the previous section and obtain predictions of
our model for the  differential branching fractions, 
forward-backward asymmetry and longitudinal polarization fraction. They are plotted
in Figs.~\ref{fig:brbseta}--\ref{fig:brbk}. By solid lines we
show results for the nonresonant branching fractions, where long-distance
contributions (\ref{eq:ybw}) of the charmonium resonances to the
coefficient $c_9^{\rm eff}$ are neglected. Plots given by the dashed lines contain
such resonant contributions. For
decays with the muon pair two largest peaks  correspond to
the contributions coming from the lowest vector charmonium states
$J/\psi$ and $\psi(2S)$, since they are narrow. The region of these resonance
peaks is excluded in experimental studies of these
decays. Contributions in the low recoil region originating from the higher vector
charmonium states, which are above the open charm threshold, are
significantly less pronounced.   Note that very recently the LHCb
Collaboration observed a charmonium resonance in the similar rare
decay $B\to K\mu^+\mu^-$ at low recoil \cite{lhcbbk}. Experimental
data are available only for $B_s \to \varphi \mu^+\mu^-$ decays. In
Figs.~\ref{fig:brbsphi}, \ref{fig:brbk} and in Tables~\ref{brcexp}, \ref{flcexp}  we
confront our predictions  for differential branching fractions, $dBr/dq^2$, and the 
longitudinal polarization fraction, $F_L$, with experimental data from
PDG (CDF) \cite{pdg} and recent LHCb \cite{lhcbbr} data. The LHCb
values for the differential branching fractions in most $q^2$ bins are  lower
than the CDF ones, but experimental error bars are rather large. Our
predictions lie just  in between these experimental
measurements. For the $\varphi$ longitudinal polarization fraction, $F_L$,
only LHCb data are available which agree with our results within uncertainties.  

\begin{table}
\caption{Comparison of our predictions for the 
  branching fractions of the rare semileptonic $B_s\to\varphi \mu^+\mu^- $ decays
  in several bins of $q^2$
  with experimental data (in $10^{-7}$). }
\label{brcexp}
\begin{ruledtabular}
\begin{tabular}{ccccc}
 $q^2$ bin (GeV$^2$)& \multicolumn{2}{c}{Theory} &  \multicolumn{2}{c}{Experiment} \\
\cline{2-3}  
\cline{4-5}
&nonresonant&resonant&PDG \cite{pdg} &LHCb \cite{lhcbbr}\\
\hline
$0.10<q^2<2.00$& $1.4\pm0.2$& $1.4\pm0.2$& $2.8\pm1.3$& $0.944\pm0.241$\\
$2.00<q^2<4.30$& $0.73\pm0.08$& $0.80\pm0.09$& $0.6\pm0.6$&$0.529\pm0.191$\\
$4.30<q^2<8.68$ & $1.8\pm0.2$& $2.4\pm0.3$& $1.3\pm0.9$&$1.38\pm0.29$ \\
$10.09<q^2<12.86$& $1.9\pm0.2$& $1.6\pm0.2$& $3.0\pm1.3$&$1.18\pm0.26$ \\
$14.18<q^2<16.00$ & $1.4\pm0.2$& $1.2\pm0.2$& $1.9\pm0.9$&$0.759\pm0.209$ \\
$16.00<q^2$& $1.7\pm0.2$& $1.5\pm0.2$& $2.3\pm1.1$& $1.06\pm0.26$\\
$1.00<q^2<6.00$& $1.7\pm0.2$& $1.9\pm0.2$& $1.1\pm0.9$&$1.14\pm0.28$ \\
$0.10<q^2<4.30$& $2.1\pm0.2$& $2.2\pm0.2$& $3.3\pm1.5$& $1.47\pm0.23$\\
\end{tabular}
\end{ruledtabular}
\end{table}

\begin{table}
\caption{Comparison of our predictions for the 
  longitudinal polarization fraction $F_L$  of the rare semileptonic $B_s\to\varphi \mu^+\mu^- $ decays
  in several bins of $q^2$
  with experimental data. }
\label{flcexp}
\begin{ruledtabular}
\begin{tabular}{cccc}
 $q^2$ bin (GeV$^2$)& \multicolumn{2}{c}{Theory} &  {Experiment} \\
\cline{2-3}  
&nonresonant&resonant&LHCb \cite{lhcbbr}\\
\hline
$0.10<q^2<2.00$& $0.50\pm0.05$& $0.56\pm0.06$& $0.37^{+0.20}_{-0.18}$\\
$2.00<q^2<4.30$& $0.73\pm0.07$& $0.85\pm0.09$&$0.53^{+0.27}_{-0.25}$\\
$4.30<q^2<8.68$ & $0.54\pm0.05$& $0.77\pm0.08$&$0.81^{+0.12}_{-0.14}$  \\
$10.09<q^2<12.86$& $0.40\pm0.04$& $0.35\pm0.04$&$0.33^{+0.15}_{-0.13}$ \\
$14.18<q^2<16.00$ & $0.34\pm0.03$& $0.29\pm0.03$&$0.34^{+0.19}_{-0.18}$ \\
$16.00<q^2<19.00$& $0.31\pm0.03$& $0.28\pm0.03$&$0.16^{+0.18}_{-0.12}$\\
$1.00<q^2<6.00$& $0.68\pm0.07$& $0.80\pm0.08$& $0.56^{+0.19}_{-0.18}$ \\
\end{tabular}
\end{ruledtabular}
\end{table}

\begin{sidewaystable}
\caption{Comparison of theoretical predictions for the nonresonant
  branching fractions of the rare semileptonic $B_s$ decays and
  available experimental data (in $10^{-7}$). }
\label{brbk}
\begin{ruledtabular}
\begin{tabular}{lcccccccccc}
 Decay& This &\multicolumn{3}{c} {\cite{ccf}}& \cite{gl} &\cite{akf}& 
 \cite{c}&  \cite{wzz}&\multicolumn{2}{c}{Experiment} \\
\cline{3-5} \cline{10-11}
&paper&A&B&C&&&&& PDG \cite{pdg} & LHCb \cite{lhcbbr}\\
\hline
$B_s\to \eta \mu^+\mu^-$ & $3.6\pm0.4$ &$1.2\pm0.3$ &$2.6\pm0.7$&$3.4\pm1.8$&
3.12 & $2.30\pm0.97$ & 2.4 &
$1.2\pm0.12$\\
$B_s\to \eta \tau^+\tau^-$ & $0.87\pm0.09$&$0.30\pm0.05$
&$0.80\pm0.15$&$1.0\pm0.55$& 0.67 & $0.373\pm0.156$ & 0.58 &
$0.34\pm0.04$\\
$B_s\to \eta \nu\bar \nu$ & $23.1\pm2.3$ &$9.5\pm2$ &$22\pm7$&$29\pm15$&
21.7 & $13.5\pm5.6$ & 17 &\\
$B_s\to \eta' \mu^+\mu^-$ & $3.1\pm0.3$ &$1.1\pm0.3$ &$2.2\pm0.6$&$2.8\pm1.5$&
3.42 & $2.24\pm0.94$ & 1.8 &\\
$B_s\to \eta' \tau^+\tau^-$ & $0.37\pm0.04$&$0.155\pm0.03$
&$0.385\pm0.075$&$0.47\pm0.25$& 0.43 & $0.280\pm0.118$ & 0.26 &\\
$B_s\to \eta' \nu\bar \nu$ & $19.7\pm2.0$ &$9\pm2$ &$19\pm5$&$24\pm13$&
23.8 & $13.3\pm5.5$ & 13 &\\
$B_s\to \varphi \mu^+\mu^-$ & $11.1\pm1.1$ & &&&16.4 &  &  &
$11.8\pm1.1$& $12.3^{+4.0}_{-3.4}$& $7.07^{+0.97}_{-0.94}$\\
$B_s\to \varphi \tau^+\tau^-$& $1.5\pm0.2$ & &&&1.51 &  &  &
$1.23\pm0.11$\\ 
$B_s\to \varphi \nu\bar \nu$& $79.6\pm8.0$ & &&&116.5 &  &  & & $<54000$  &\\ 
$B_s\to K \mu^+\mu^-$ & $0.22\pm0.02$ & &&& &  &  0.14&
$0.199\pm0.021$\\
$B_s\to K \tau^+\tau^-$ & $0.055\pm0.006$ & &&& &  &  0.03&
$0.074\pm0.007$\\
$B_s\to K \nu\bar\nu$ & $1.41\pm0.14$ & &&& &  &  1.01&\\
$B_s\to K^* \mu^+\mu^-$ & $0.42\pm0.04$ & &&& &  &  &
$0.38\pm0.03$\\
$B_s\to K^* \tau^+\tau^-$ & $0.075\pm0.008$ & &&& &  &  &
$0.050\pm0.004$\\
$B_s\to K^* \nu\bar\nu$ & $3.0\pm0.3$ & &&& &  &  &&
\end{tabular}
\end{ruledtabular}
\end{sidewaystable}

Now we integrate the differential branching fraction over $q^2$ and
get the results for the total branching fractions. For the evaluation of
the branching fractions of the rare $B_s\to K$ decays, governed by the
$b\to d$ weak current, we use the form factors previously calculated
in our model in Ref.~\cite{bsld}. 
In Table~\ref{brbk} we present our predictions for the nonresonant
branching fractions of the rare semileptonic $B_s$ decays and compare
them with previous calculations \cite{ccf,gl,akf,c,wzz} and available
experimental data \cite{pdg,lhcbbr}. In Ref.~\cite{ccf} three 
sets of form factors based on different versions of sum rules were
considered. Set A uses short-distance QCD 
sum rules. Set B is based on light-cone QCD sum rules, while set C
arises from light-cone QCD sum rules within the soft collinear
effective theory. The authors of Ref.~\cite{gl} employ the light front
and constituent quark models for the evaluation of the rare decay
branching fractions. Three-point QCD sum rules are used for the analysis
of the rare semileptonic $B_s$ decays into $\eta(\eta')$ and lepton
pair in Ref.~\cite{akf}. In Ref.~\cite{c} calculations are based on
the light-front quark model, while light-cone sum rules in the
framework of heavy quark effective field theory are applied in
Ref.~\cite{wzz}. The analysis of the predictions given in
Table~\ref{brbk} indicate that  these significantly different
approaches give close values of order  $10^{-7}$ for the rare
semileptonic $B_s\to \varphi(\eta^{(')})l^+l^-$ decay branching
fractions and of order  $10^{-8}$ for $B_s\to K^{(*)}l^+l^-$ decays.   
Experimental data are available for the branching
fraction of the $B_s\to\varphi\mu^+\mu^-$ decay only. As we see from
the table all 
theoretical predictions are well consistent with each other and
experimental data for the $B_s\to\varphi\mu^+\mu^-$ decay from PDG
\cite{pdg}. Note that very recently the LHCb Collaboration
\cite{lhcbbr} also reported measurement of this decay branching
fraction which is somewhat lower than previous measurements. Our
prediction is consistent with the latter value within 2$\sigma$.

\section{$B_s\to \eta(\varphi) \nu\bar\nu$ decays} 
\label{sec:bsnu}

The differential decay rate for the $B_s\to \eta(\varphi)\nu\bar\nu$
decay is given
by \cite{fgikl}
\begin{equation}
  \label{eq:dgnunu}
  \frac{d\Gamma(B_s\to \eta(\varphi)\nu\bar\nu)}{dq^2}=3\frac{G_F^2}{(2\pi)^3}
  \left(\frac{\alpha |V_{ts}^*V_{tb}|}{2\pi}\right)^2
  \frac{\lambda^{1/2}q^2}{24M_{B_s}^3}
  H^{(\nu)}H^{\dag(\nu)},
\end{equation}
where the factor  3 originates from the sum over neutrino flavours,
\[  H^{(\nu)}H^{\dag(\nu)}\equiv
H^{(\nu)}_+H^{\dag(\nu)}_++H^{(\nu)}_-H^{\dag(\nu)}_-+H^{(\nu)}_0H^{\dag(\nu)}_0,\]
and the helicty amplitudes $H^{(\nu)}_m$ are expressed through form
factors by the following relations.

(a) $B\to \eta^{(')}$ transition: 
\begin{eqnarray}
  \label{eq:hpnu}
  H^{(\nu)}_\pm&=&0,\cr
H^{(\nu)}_0&=&\frac{\lambda^{1/2}}{\sqrt{q^2}} C_L^\nu f_+(q^2).
\end{eqnarray}

(b) $B\to \varphi$ transition:
\begin{eqnarray}
  \label{eq:hpnuv}
 H^{(\nu)}_\pm&=&C_L^\nu\left[-(M_{B_s}+M_\varphi)A_1(q^2)\pm \lambda^{1/2}
\frac{V(q^2)}{M_{B_s}+M_\varphi}\right],\cr
H^{(\nu)}_0&=&-\frac{1}{2M_\varphi\sqrt{q^2}}C_L^\nu\left[(M_{B_s}^2-M_\varphi^2-q^2)
    (M_{B_s}+M_\varphi)A_1(q^2)-\frac{\lambda}{M_{B_s}+M_\varphi}A_2(q^2)\right].
\end{eqnarray}
Here $$C_L^\nu=-X(x_t)/\sin^2\theta_W, \qquad x_t=m_t^2/m_W^2,$$
$\theta_W$ is the Weinberg angle, and the function $X(x_t)$ at the
leading-order in QCD has the form
\[X(x_t)=\frac{x}8\left(\frac{2+x}{x-1}+\frac{3x-6}{(x-1)^2}\ln x
\right),\]
while the next-to-leading order expressions are given in
Ref.~\cite{mu}.

Substituting the experimental values for the top ($m_t$) and
$W$-boson ($m_W$) masses one gets \cite{absw}
\begin{equation}
  \label{eq:cl}
  C^\nu_L=-6.38\pm0.06,
\end{equation}
where the error is dominated by the top quark mass uncertainty. In the
following calculations we use the central value of $C^\nu_L$.

The differential  longitudinal polarization fraction $F_L$ of the $\varphi$ meson is defined
similar to Eq.~(\ref{eq:fl})
\begin{equation}
  \label{eq:flnu}
  F_L=\frac{ H^{(\nu)}_0H^{\dag(\nu)}_0}{ H^{(\nu)}H^{\dag(\nu)}}.
\end{equation}

\begin{figure}
  \centering
 \includegraphics[width=7.8cm]{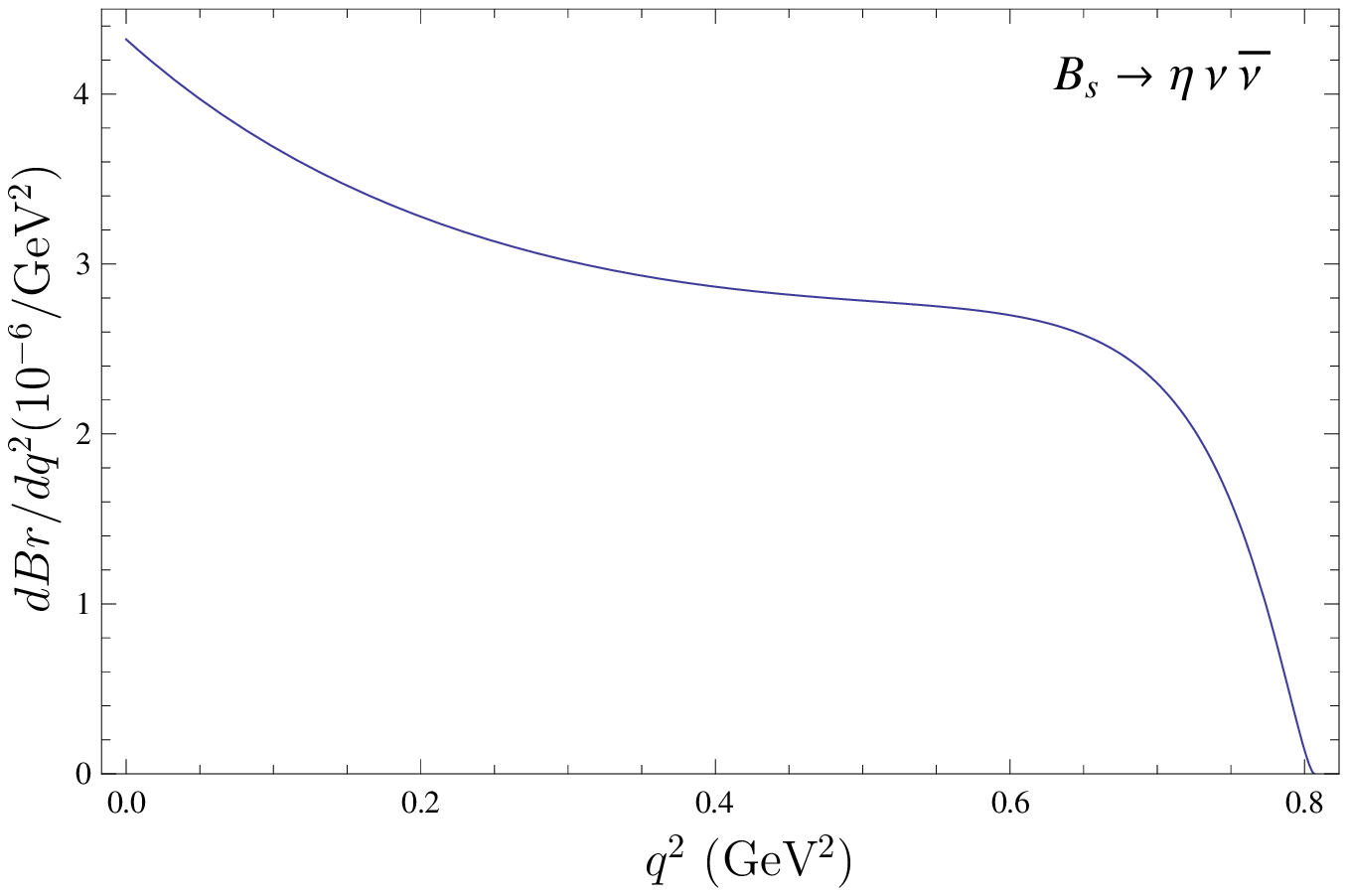} \ \
 \   \includegraphics[width=7.8cm]{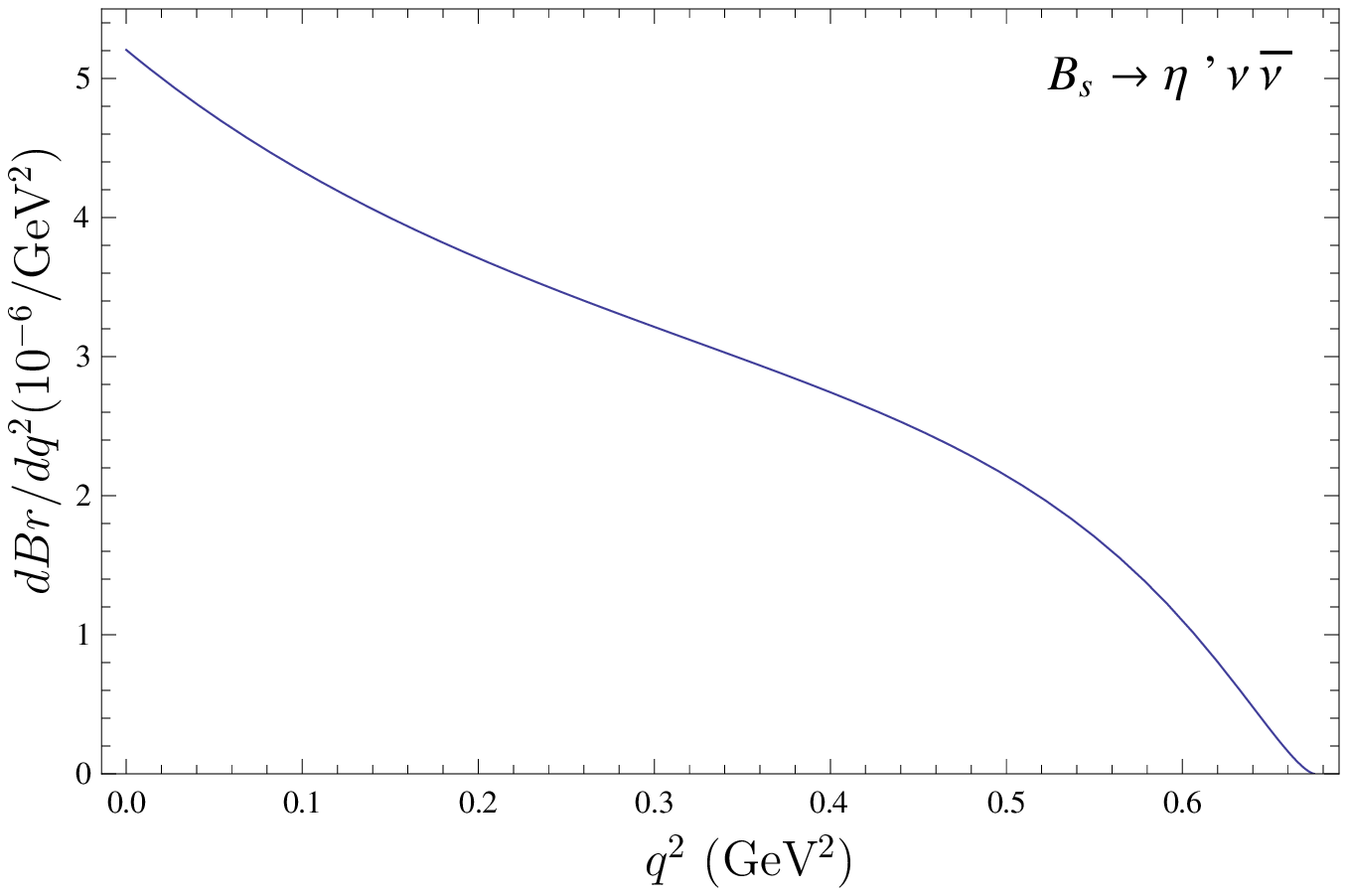}
\caption{Theoretical predictions for the differential
    branching fractions $d Br(B_s \to \eta^{(')}\nu\bar\nu)/d
    q^2$. }
  \label{fig:brbsnu}
\end{figure}
\begin{figure}
  \centering
 \includegraphics[width=8cm]{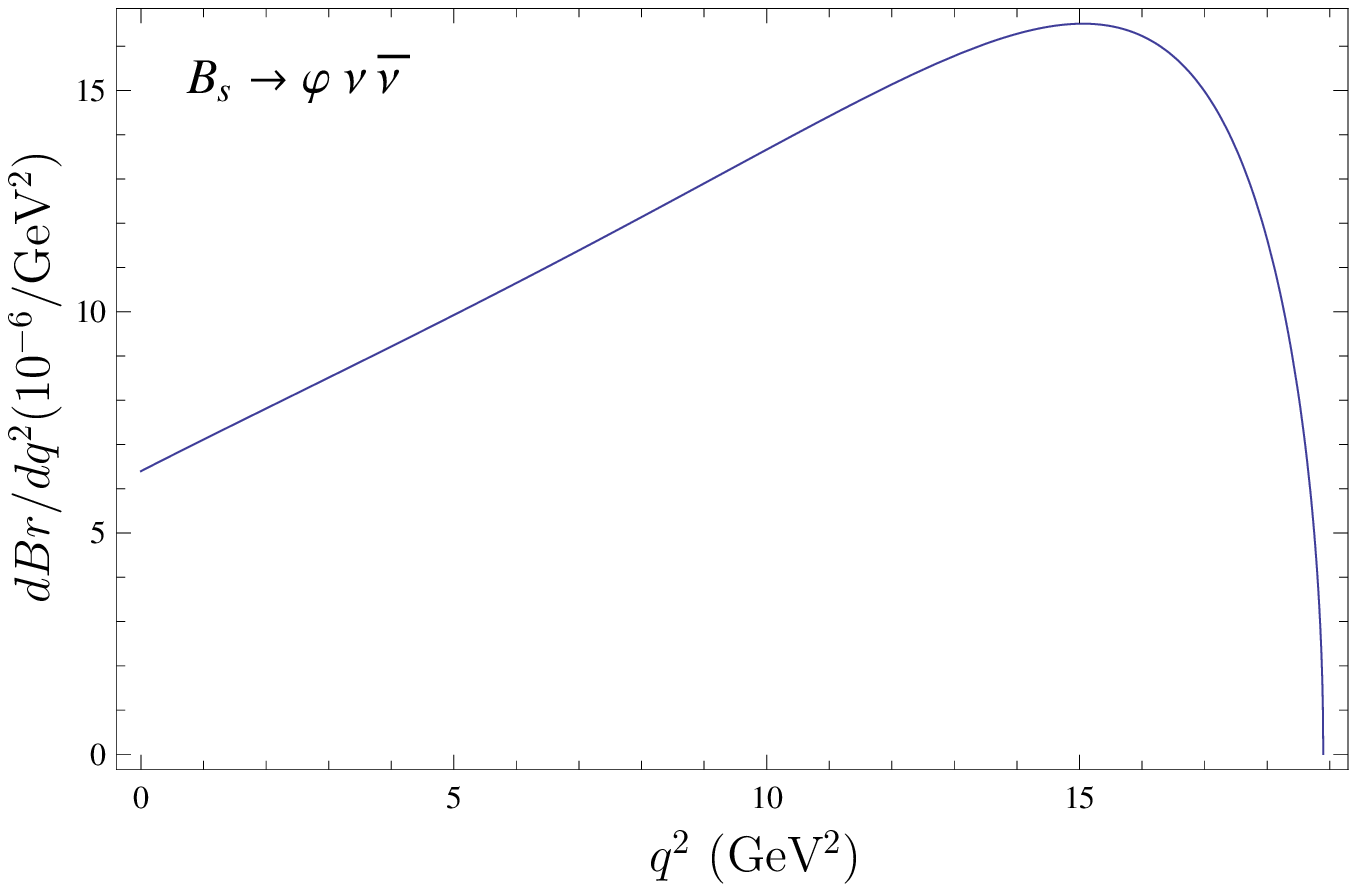}\ \ \  \includegraphics[width=8cm]{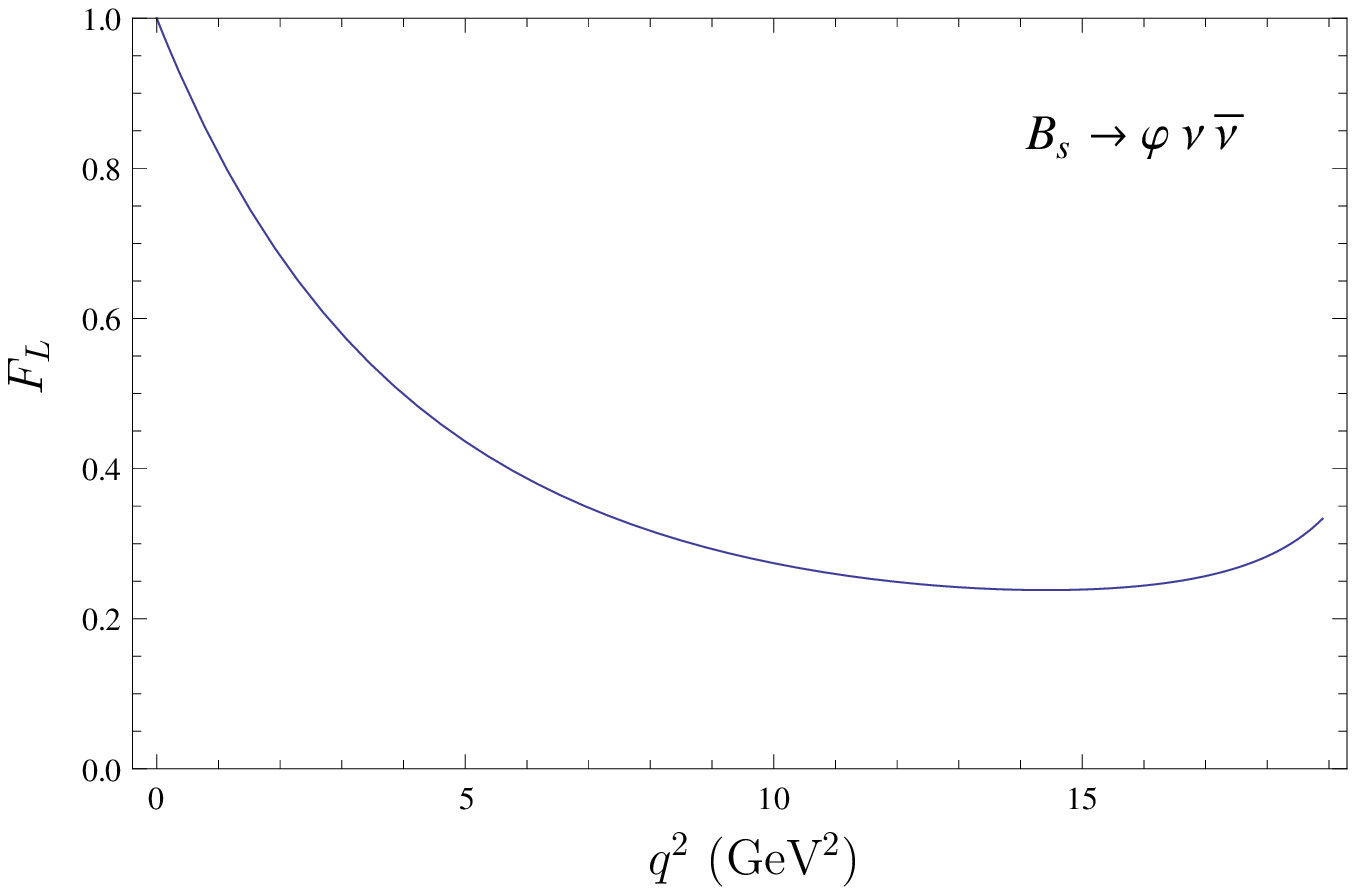}
  \caption{Theoretical predictions for the differential
    branching fractions $d Br(B_s \to \varphi \nu \bar\nu)/d
    q^2$ and longitudinal polarization fraction $F_L$ of the $\varphi$ meson. }
  \label{fig:brbsphinu}
\end{figure}

Now we substitute the rare decay form factors calculated in our model
into the above expressions for the branching fractions (\ref{eq:dgnunu}) and the
longitudinal polarization fraction (\ref{eq:flnu}). The resulting predictions for the differential branching fractions of the $B_s\to
\eta(\eta') \nu\bar\nu$ decays are plotted in Fig.~\ref{fig:brbsnu}. In
Fig.~\ref{fig:brbsphinu} we present the corresponding  differential branching
fraction and the $\varphi$ longitudinal polarization fraction ($F_L$)
for the $B_s\to \varphi \nu\bar\nu$ decay. 

We give our results for the branching fractions of the rare $B_s$
decays $B_s\to \eta(\varphi) \nu\bar\nu$ and $B_s\to K^{(*)}
\nu\bar\nu$ in Table~\ref{brbk} in comparison with previous
calculations \cite{ccf,gl,akf,c}. Again we find a reasonable agreement
between  predictions in significantly different approaches. The
obtained branching fractions are of order  $10^{-6}$ for the $B_s\to
\eta(\varphi) \nu\bar\nu$ and  $10^{-7}$ for $B_s\to K^{(*)}
\nu\bar\nu$ decay branching fractions. At present,  
only rather loose experimental upper bound (of order  $10^{-3}$) is available 
for the $B_s\to \varphi \nu\bar \nu$ decay branching fraction. Of
course, all predictions are well below this limit.

\section{Rare radiative $B_s$ decays}
\label{sec:rrad}

The exclusive rare radiative decay rate $B_s\to \varphi\gamma$ for
the emission of a real photon ($k^2=0$) is determined by the form factor
$T_{1}(0)$ and  is given by
\begin{equation}
\label{drate1}
\Gamma(B_s\to \varphi\gamma)=
\frac{\alpha }{32\pi^4} G_F^2m_b^2M_{B_s}^3|V_{tb}V_{ts}|^2
|c_7^{\rm eff}(m_b)|^2 |T_{1}(0)|^2 
\left(1-\frac{M_{\varphi}^2}{M_{B_s}^2}
 \right)^3\left( 1+\frac{M_{\varphi}^2}{M_{B_s}^2}\right).
\end{equation}

\begin{table}
\caption{Comparison of predictions for the 
  branching fractions of the rare radiative decays
  with experimental data. }
\label{brrare}
\begin{ruledtabular}
\begin{tabular}{ccccc}
& \multicolumn{2}{c}{Theory}&  \multicolumn{2}{c}{Experiment} \\
\cline{2-3}  \cline{4-5}
 Decay& This paper& \cite{ap} &PDG \cite{pdg} &LHCb \cite{lhcbrad}\\
\hline
$Br(B^0\to K^{*0}\gamma)\times 10^{5}$& $4.3\pm0.4$& $4.3\pm1.4$& $4.33\pm0.15$& \\
$Br(B_s\to \varphi\gamma)\times10^{5}$& $3.8\pm0.4$& $4.3\pm1.4$&
$5.7^{+2.2}_{-1.9}$& $3.5\pm0.4$\\
$\frac{Br(B^0\to K^{*0}\gamma)}{Br(B_s\to \varphi\gamma)}$& $1.14\pm0.12$& $1.0\pm0.2$&
$0.7\pm0.3$& $1.23\pm0.12$\\
$Br(B_s\to K^{*0}\gamma)\times 10^{5}$& $0.13\pm0.02$&
\end{tabular}
\end{ruledtabular}
\end{table}

To evaluate the rare radiative $B_s\to \varphi\gamma$ decay rate
we substitute the value of the form factor $T_{1}(0)$ from
Table~\ref{hff} in  the expression (\ref{drate1}). The result is given
in Table~\ref{brrare}. There we also show our previous prediction for
the $Br(B^0\to K^{*0}\gamma)$ \cite{brad}. In this table we confront
our results with the values obtained in the framework of the soft
collinear effective theory at NNLO \cite{ap} and available
experimental data \cite{pdg,lhcbrad}. We see that both theoretical
predictions agree well with experimental values. We
also compare results for the ratio of the rare radiative branching
fractions of $B$ and $B_s$ decays. This ratio was recently measured
with  improved precision by
the LHCb Collaboration \cite{lhcbrad}. The central value was found
to be significantly larger than the previous one \cite{pdg}, but the
errors are still large. Note that our result is more close to the
LHCb value but is consistent with both experimental values. In
Table~\ref{brrare} we also give our prediction for the still
unmeasured CKM suppressed $B_s\to
K^{*0}\gamma$ decay. 

\section{Rare nonleptonic $B_s$ decays}
\label{sec:nonl}

Next we use the calculated form factors for the evaluation of the two-body
nonleptonic decays of $B_s$ mesons governed by the rare weak $b\to s$
($b\to d$)
transition. Following our previous calculations of the nonleptonic
$B_s$ decays \cite{bsdecay,bsld} we use the factorization approximation. As
a result the complicated nonleptonic decay amplitude reduces to
the product of the matrix element of the weak current between the $B_s$ meson
and the final $\eta^{(')}$ or $\varphi$ meson with the matrix element of the
weak current between the second meson and vacuum. For example the rare
nonleptonic decay
amplitude for $B_s\to \eta^{(')}\psi$ ($\psi$ denotes the $c\bar
c$ meson) can be approximated by the product of the one-particle matrix elements     
\begin{equation}
  \label{eq:fact} 
\langle  \eta^{(')}\psi|H_{\rm eff}|B_s\rangle \approx \frac{G_F}{\sqrt{2}}
V_{cb}^*V_{cs} a_2^{\rm eff}\langle  \eta^{(')}|\bar s\gamma^\mu(1-\gamma^5)b|B_s\rangle \langle \psi|(\bar c\gamma_\mu(1-\gamma^5)c|0\rangle,
\end{equation}
with the effective coefficient
\begin{equation}
  \label{eq:aeff}
a_2^{\rm
  eff}=a_2-\frac{V_{tb}^*V_{ts}}{V_{cb}^*V_{cs}}[a_3+a_5+a_{7}+a_9]\approx
a_2+a_3+a_5+a_{7}+a_9, 
\end{equation}
where terms in square brackets result from the
contributions of  penguin diagrams. In the right-hand-side of Eq.~(\ref{eq:aeff}) we used the
approximate relation, $V_{tb}^*V_{ts}\approx-V_{cb}^*V_{cs}$, following
from the unitarity of the CKM matrix. 
The quantities $a_{2n-1}=c_{2n-1}+c_{2n}/N_c$ and
$a_{2n}=c_{2n}+c_{2n-1}/N_c$ ($n=1,2\dots$ and $N_c$ is the number of colors) are combinations of
the Wilson coefficients $c_i$. The
similar expressions can be obtained for other nonleptonic decays
considered in this paper.

The matrix element of the weak current
$J^W_\mu$ between meson states is expressed through decay form factors
calculated in Sec.~\ref{sec:ffbsphi}, while the matrix elements
between vacuum and a final  
pseudoscalar ($P$), vector ($V$) or axial vector ($A$) meson are parametrized by the decay
constants $f_{P,V,A}$
\begin{eqnarray}
\langle P|\bar q_1 \gamma^\mu\gamma_5 q_2|0\rangle&=&if_Pp^\mu_P, \cr
\langle V|\bar q_1\gamma_\mu q_2|0\rangle&=&\epsilon_\mu M_Vf_V,\cr
\langle A|\bar q_1\gamma^\mu\gamma_\mu q_2|0\rangle&=&\epsilon_\mu M_Af_A.
\end{eqnarray}
For the calculations we use the following values of the decay
constants: $f_K=0.156$~GeV, $f_{K^*}=0.214$~GeV, $f_\varphi=0.231$~GeV,
$f_{J/\psi}=0.415$~GeV,  $f_{\chi_{c1}}=0.161$~GeV, and the central
values of the CKM
matrix elements: $|V_{cs}|=0.973$,  $|V_{cb}|=0.039$,
$|V_{td}|=0.0087$, $|V_{ts}|=0.0404$,  $|V_{tb}|=0.999$ \cite{pdg}. 
    
\begin{table}
\caption{Comparison of theoretical predictions for the branching fractions
  of the nonleptonic $B_s$  decays with available  experimental data (in $10^{-4}$).  }
\label{compnl}
\begin{ruledtabular}
\begin{tabular}{@{}c@{}c@{}c@{$\!$}c@{}c@{}cc@{}c@{}c@{}c@{}}
& \multicolumn{5}{c}{Theory}&  \multicolumn{2}{c@{}}{Experiment} \\
\cline{2-6}  \cline{7-8}
Decay&this paper&\cite{cfw}
&\cite{ikkss,ddil}&\cite{bn}&\cite{akllsww}&PDG \cite{pdg}&
LHCb \cite{lhcbnl,lhcbnl2} \\
\hline
$B_s\to J/\psi\eta$& $3.6\pm0.6$ & $4.2\pm0.2$&$4.67$&&
&$5.1^{+1.3}_{-1.0}$  & $3.79^{+0.73}_{-0.80}$ \\
$B_s\to J/\psi\eta'$& $3.7\pm0.6$ & $4.3\pm0.2$&$4.04$&&
&$3.7^{+1.0}_{-0.9}$  & $3.42^{+0.66}_{-0.73}$ \\
$B_s\to J/\psi\varphi$& $11.3\pm1.6$ & $16.7\pm5.7$&$16$&&
&$10.9^{+2.8}_{-2.3}$  & $\! 10.5\pm1.05$ \\
$B_s\to \psi(2S)\eta$& $1.9\pm0.3$ & $3.0\pm0.2$&&  &  \\
$B_s\to \psi(2S)\eta'$& $1.6\pm0.3$ & $2.5\pm0.2$&&  &  \\
$B_s\to \psi(2S)\varphi$& $6.9\pm0.9$ & $8.3\pm2.7$&&&
&$5.7^{+1.8}_{-1.6}$  &  \\
$B_s\to \chi_{c1}\eta$& $0.56\pm0.09$ & $2.0\pm0.2$&&  & && \\
$B_s\to \chi_{c1}\eta'$& $0.51\pm0.08$ & $1.8\pm0.2$&&  & && \\
$B_s\to \chi_{c1}\varphi$& $1.95\pm0.09$ & $3.3\pm1.3$&&  &  &&\\
$B_s\to \varphi\eta$& $0.018\pm0.003$ & &&$\!\! 0.0012^{+0.0139}_{-0.0023}$ &$0.036^{+0.017}_{-0.012}$
&&  \\
$B_s\to \varphi\eta'$& $0.021\pm0.003$ & & &$\!\! 0.0005^{+0.0118}_{-0.0019}$ &$0.0019^{+0.0020}_{-0.0013}$
&&  \\
$B_s\to \varphi\varphi$& $0.22\pm0.03$ & &&$0.218^{+0.304}_{-0.170}$ &$0.353^{+0.187}_{-0.123}$
&$0.19^{+0.06}_{-0.05}$  &  \\
$B_s\to J/\psi K$& $0.25\pm0.05$ & &&&&$0.36\pm0.08$  &  \\
$B_s\to J/\psi K^*$& $0.57\pm0.09$ & &&&&$0.9\pm0.4$  &  \\
$B_s\to \psi(2S) K$& $0.12\pm0.02$ & &&  &  \\
$B_s\to \psi(2S) K^*$& $0.36\pm0.06$ & &&  &  \\
$B_s\to K^+K^-$& $0.19\pm0.03$ & &&$0.227^{+0.275}_{-0.130}$ &$0.136^{+0.086}_{-0.052}$
&$0.264\pm0.028$&  \\
$B_s\to K^+K^{*-}$& $0.27\pm0.05$ & &&$0.055^{+0.151}_{-0.047}$
&$0.047^{+0.027}_{-0.016}$ &&  \\
$B_s\to K^{*+}K^-$& $0.15\pm0.03$ & &&$0.041^{+0.096}_{-0.032}$
&$0.060^{+0.025}_{-0.020}$ &&  \\
$B_s\to K^{*+}K^{*-}$& $0.29\pm0.05$ & &&$0.091^{+0.105}_{-0.063}$
&$0.067^{+0.039}_{-0.022}$ &&  \\
$B_s\to K^0\bar K^0$& $0.15\pm0.03$ & &&$0.247^{+0.293}_{-0.140}$ &$0.156^{+0.097}_{-0.060}$
&$<0.66$&  \\
$B_s\to \bar K^0K^{*0}$& $0.24\pm0.05$ & &&$0.039^{+0.106}_{-0.035}$
&$0.073^{+0.033}_{-0.022}$ &&  \\
$B_s\to \bar K^{*0}K^0$& $0.13\pm0.03$ & &&$0.042^{+0.140}_{-0.040}$
&$0.043^{+0.023}_{-0.016}$ &&  \\
$B_s\to K^{*0}\bar K^{*0}$& $0.26\pm0.05$ & &&$0.091^{+0.113}_{-0.068}$
&$0.078^{+0.043}_{-0.027}$ &$0.28\pm0.07$&  \\
$\!\!\! B_s\to \varphi\bar K^{*0}$& $\!\!\! 0.0076\pm0.0012$ & &&$0.004^{+0.005}_{-0.003}$
&$0.0065^{+0.0033}_{-0.0023}$&$<10$ &$\!\! 0.011\pm0.003$  \\
\end{tabular}
\end{ruledtabular}
\end{table}

In Table~\ref{compnl} we compare our predictions for the rare
nonleptonic $B_s$ decays with other theoretical calculations
\cite{cfw,ikkss,ddil,bn,akllsww}  and
available experimental data \cite{pdg,lhcbnl,lhcbnl2}. The authors of
Ref.~\cite{cfw} analyze rare nonleptonic $B_s$ decays to the ground or
excited charmonium states and light mesons using generalized
factorization and $SU(3)$ symmetry to relate such modes to
corresponding $B$ decays. Nonleptonic $B_s\to J/\psi \eta^{(')}$ and
$B_s\to J/\psi \varphi$ decays are studied in Refs.~\cite{ikkss,ddil}
in the framework of the covariant constituent quark
model. Considerations of the rare nonleptonic $B_s$ decays to light
final mesons in Ref.~\cite{bn} are based on QCD factorization, while
perturbative QCD approach is adopted in
Ref.~\cite{akllsww}. Comparison of our predictions for the rare $B_s$
decays to charmonium states and a light mesons with results of
Refs.~\cite{cfw,ikkss,ddil} shows that our model yields the central
values 30-60\% lower, especially for decays involving
$\chi_{c1}$. However results are still compatible taking into account
the uncertainties. They are also in agreement with the available
experimental data, our central values for these decays being slightly
closer to the experimental central values recently published by the
LHCb Collaboration \cite{lhcbnl}. Very recently the Belle
Collaboration \cite{bellebphi} reported a measurement of the branching fraction of the
decay $B_s\to J/\psi\varphi$ with the value
$Br(B_s\to J/\psi\varphi)=(1.25\pm0.07\pm0.08\pm0.22)\times 10^{-3}$. For the rare $B_s$ decays to the
pair of light mesons our predictions in general agree with results of
Refs.~ \cite{bn,akllsww} and experiment taking into account rather large error bars. Note that very recently the LHCb Collaboration
\cite{lhcbnl2} reported the first observation of the decay $B_s\to
\varphi\bar K^{*0}$. The obtained central value of this decay
branching fraction is found to be larger than theoretical expectations
(our result is the closest to experiment) 
but the experimental and theoretical values agree within error bars.

\begin{table}
\caption{Comparison of theoretical predictions for the ratios of branching fractions
  of the nonleptonic $B_s$  decays with experimental data.  }
\label{compnlr}
\begin{ruledtabular}
\begin{tabular}{cccccc}
& \multicolumn{3}{c}{Theory}&  \multicolumn{2}{c}{Experiment} \\
\cline{2-4}  \cline{5-6}
Ratio&this paper&\cite{cfw}
&\cite{ikkss,ddil}&PDG \cite{pdg}&
LHCb \cite{lhcbnl} \\
\hline
$\frac{Br(B_s\to J/\psi\eta')}{Br(B_s\to J/\psi\eta)}$& $1.03\pm0.10$ & $1.02\pm0.05$&$0.87$
&$0.73\pm0.14$  & $0.90^{+0.11}_{-0.09}$ \\
$\frac{Br(B_s\to \psi(2S)\eta)}{Br(B_s\to J/\psi\eta)}$& $0.53\pm0.05$
& $0.71\pm 0.24$&
& & $0.83\pm 0.19$ \\
$\frac{Br(B_s\to \psi(2S)\varphi)}{Br(B_s\to J/\psi\varphi)}$& $0.61\pm0.06$ & $0.50\pm0.17$&
&$0.53\pm 0.10$ &  \\
$\frac{Br(B_s\to \chi_{c1}\varphi)}{Br(B_s\to J/\psi\varphi)}\times 10^2$& $17.3\pm1.7$ & $19.8\pm6.7$&
& & $18.9\pm 2.4$ \\
$\frac{Br(B_s\to \varphi\mu^+\mu^-)}{Br(B_s\to J/\psi\varphi)}\times 10^4$& $10.3\pm1.0$ && & $11.3\pm2.0$&
 $6.74\pm 0.63$ \\
\end{tabular}
\end{ruledtabular}
\end{table}

In Table~\ref{compnlr} we present comparison of predictions for the ratios of branching
fractions of the rare nonleptonic $B_s$ decays to the charmonium state
and a light meson with experimental data which have smaller error
bars. In general, good agreement of theory with experiment is
observed. In this table we also give the ratio of the branching
fractions of the rare
semileptonic $B_s\to \varphi\mu^+\mu^-$ and nonleptonic $B_s\to
J/\psi\varphi$ decays. Our prediction is in accord with the PDG value
\cite{pdg}, but almost a factor 1.5 larger than the recent LHCb
value. This is the consequence of the significantly lower LHCb
value for $Br(B_s\to \varphi\mu^+\mu^-)$ as it was already mentioned
in Sec.~\ref{sec:rard}.  

\section{Conclusions}
\label{sec:concl}

The form factors parametrizing the transition matrix element of the flavour
changing neutral current, governed by $b\to s$ quark transition,
between the $B_s$ and  light ($\eta(\eta')$ or $\varphi$) mesons were
calculated on the basis of the relativistic quark model with the
QCD-motivated quark-antiquark  interaction potential. All relativistic effects,
including boosts of the meson wave functions and contributions of the
intermediate negative-energy states, were consistently taken into
account. The main advantage of the adopted approach consists in 
 that it allows the determination of the momentum transfer
dependence of the form factors in the whole accessible kinematical
range. Therefore no additional assumptions and ad hoc extrapolations
are needed for the description of the rare weak decay
processes which have rather broad kinematical range. This
significantly improves the reliability of the obtained results.

The calculated form factors were used while considering  the rare
semileptonic, radiative and nonleptonic $B_s$ decays. The differential
and total decay branching fractions as well as asymmetry and
polarization parameters were evaluated. The obtained results were
confronted with previous investigations based on significantly different
theoretical approaches and available experimental data. Good agreement
of our predictions with measured values is observed.    

\acknowledgments
The authors are grateful to A. Ali, D. Ebert, C. Hambrock,
M.~A.~Ivanov, V. A. Matveev and V. I. Savrin  
for  useful discussions.
This work was supported in part by the {\it Russian
Foundation for Basic Research} under Grant No.12-02-00053-a.

\end{document}